\documentclass[prd,preprintnumbers,superscriptaddress,nofootinbib,amssymb,aps,twocolumn,letterpaper,floatfix]{revtex4-1}
\usepackage[usenames]{color} 
\usepackage{graphicx}
\usepackage{amssymb}
\usepackage{amsmath}
\usepackage[normalem]{ulem}
\usepackage{xspace}
\include{parameterdefs}

\renewcommand{\Re}{\mathop{\mathrm{Re}}\nolimits}

\begin{document}

\newcommand*{\CSUF}{California State University Fullerton, Fullerton, CA 92831, USA}\affiliation{\CSUF}
\newcommand*{\caltech}{California Institute of Technology, Pasadena, CA 91109, USA}\affiliation{\caltech}
\newcommand*{\OSK}{Institute of Laser Engineering, Osaka University, Suita, 567-0086, 
Japan}\affiliation{\OSK}
\newcommand*{\YITP}{Yukawa Institute for Theoretical Physics, Kyoto
University, Kyoto, 606-8502, Japan}\affiliation{\YITP}
\newcommand*{\uwm}{Department of Physics, University of Wisconsin--%
  Milwaukee, PO Box 413, Milwaukee, WI 53201, USA}\affiliation{\uwm}
\newcommand*{\JILA}{JILA, University of Colorado and National Institute of Standards and Technology,
  440 UCB, Boulder, CO 80309, USA}\affiliation{\JILA}
\newcommand*{\Jena}{Theoretisch-Physikalisches Institut, Friedrich Schiller Universit{\"a}t Jena, Max-Wien-Platz 1, 07743 Jena, Germany}\affiliation{\Jena}
\newcommand*{\AEI}{Max-Planck-Institut f\"ur Gravitationsphysik, Albert-Einstein-Institut, Am M\"uhlenberg 1, D-14476 Golm, Germany}\affiliation{\AEI}

\newcommand{\programname}[1]{\texttt{#1}}
\newcommand{\sacra}{\programname{SACRA}\xspace}
\newcommand{\whisky}{\programname{Whisky}\xspace}

\newcommand{\cf}{\textrm{cf.}~}
\newcommand{\ie}{\textrm{i.e.}~}
\newcommand{\eg}{\textrm{e.g.}~}
\newcommand{\ms}{{\rm ms}}
\newcommand{\km}{{\rm km}}
\newcommand{\hz}{{\rm Hz}}
\newcommand{\khz}{{\rm kHz}}
\newcommand{\mpc}{{\rm Mpc}}

\newlength{\digitwidth}
\newlength{\signwidth}
\settowidth{\digitwidth}{0}
\settowidth{\signwidth}{+}

\makeatletter
\def\hlinewd#1{%
\noalign{\ifnum0=`}\fi\hrule \@height #1 %
\futurelet\reserved@a\@xhline}
\makeatother

\newcommand{\raisedrule}[2][0em]{\leaders\hbox{\rule[#1]{1pt}{#2}}\hfill}

\title{Matter effects on binary neutron star waveforms }

\author{Jocelyn~S.~Read}\affiliation{\CSUF}\affiliation{\caltech}
\author{Luca~Baiotti}\affiliation{\OSK}\affiliation{\YITP}
\author{Jolien~D.~E.~Creighton}\affiliation{\uwm}
\author{John~L.~Friedman}\affiliation{\uwm}
\author{Bruno~Giacomazzo}\affiliation{\JILA}
\author{Koutarou~Kyutoku}\affiliation{\uwm}
\author{Charalampos~Markakis}\affiliation{\Jena}\affiliation{School of Mathematics, University of Southampton, Southampton, SO17 1BJ, United
Kingdom} 
\author{Luciano~Rezzolla}\affiliation{\AEI} 
\author{Masaru~Shibata}\affiliation{\YITP}
\author{Keisuke~Taniguchi}\affiliation{Graduate School of Arts and Sciences, University of Tokyo, Komaba, Meguro, Tokyo, 153-8902, Japan}

\begin{abstract}
Using an extended set of equations of state and a 
multiple-group multiple-code collaborative effort to generate waveforms, we
improve numerical-relativity-based data-analysis estimates of the
measurability of matter effects in neutron-star binaries. 
We vary two parameters 
of a parameterized piecewise-polytropic equation of state (EOS) to analyze the 
measurability of EOS properties, via a parameter $\Lambda$ that characterizes
the quadrupole deformability of an isolated neutron star.  We find that, to
within the accuracy of the simulations, the departure of the waveform from
point-particle (or spinless double black-hole binary) inspiral
increases monotonically with $\Lambda$, and changes in the EOS 
that did not
change $\Lambda$ are not measurable.

We estimate with two methods the minimal and expected measurability of
$\Lambda$ in second- and third-generation gravitational-wave detectors. The
first estimate, using numerical waveforms alone, shows two EOS which vary in
radius by $1.3$\,km are distinguishable in mergers at 100\,Mpc.  The second
estimate relies on the construction of hybrid waveforms by matching to
post-Newtonian inspiral, and estimates that the same EOS are distinguishable in
mergers at 300\,\mpc. We calculate systematic errors arising from numerical
uncertainties and hybrid construction, and we estimate the frequency at which
such effects would interfere with template-based searches.  \end{abstract}
\maketitle 

\section{Introduction}

Substantial uncertainty remains in the equation of state (EOS) of 
cold matter above nuclear density. 
While recent analyses of X-ray bursts and 
thermal emission from quiescent low-mass X-ray binaries
\cite{Steiner2010,guver11,ozel09,ozel10a,ozel10b,suleimanov10,Zamfir2012} constrain simultaneously the mass and radius of neutron stars in X-ray
binaries, placing limits on allowed EOS, such measurements depend on burst and
atmosphere models. In contrast, observations of gravitational waves from binary 
inspiral provide a {\em model-independent} way to simultaneously 
measure the mass 
and radius of neutron stars in double neutron-star and black-hole neutron-star
binaries. 

The detection of a gravitational wave from an inspiraling binary will
determine mass parameters from the early inspiral~\cite{CutlerFlanagan}.
Strong signals may also constrain additional parameters that characterize the
EOS\@.  For widely separated neutron-star pairs, EOS effects will be minuscule;
however, binary systems drawn together by the loss of orbital angular momentum
to gravitational radiation will exhibit increasing tidal interactions through
the late stage of binary inspiral, up to tidal disruption or merger. The
effects of tidal interactions imprint an EOS signature on the gravitational
waveform of the merger of the neutron stars.

The rate of binary neutron star mergers is uncertain, but it is reasonable to
expect that Advanced LIGO\cite{AdvancedLIGO} will detect several events per
year~\cite{Abadie:2010cf}. In fact, over several years of operation, there
appears to be a good chance that a strong signal, with signal-to-noise ratio
(SNR) above 30, will be detected.

Neutron-star pairs in binary systems produce mutual tidal stresses that deform
the metric around the stars in a manner prescribed by the EOS,
via a parameter we refer to as the tidal deformability, $\Lambda$, defined in Eq.~(\ref{eq:Lambda}) below. This
parameter describes the degree to which a local metric suffers quadrupolar
deformations when in the tidal field of a companion, and scales as the fifth power of the neutron-star
radius, $R^5$.  The tidal interaction between two stars in a binary system
alters both the binding energy of the system and the gravitational-wave energy
flux~\cite{Flanagan2008,Damour2010,Vines:2011ud}, and in turn changes the phase
evolution of the gravitational waveform. When the stars are sufficiently far
apart, the phase evolution may be obtained from a detailed balance of energy
through a sequence of orbits.  This approach describes the secular evolution of
the binary system orbit under energy loss to gravitational radiation (and
distortion of the companions) and is valid while the evolution is slow and the
motions are not too relativistic. Analysis using analytic models suggests that
tidal effects may be measurable using Advanced LIGO
\cite{Flanagan2008,Hinderer2010,Damour2012}, but only if the model can be
extended to the late, high-frequency stages of inspiral. 

Large tidal effects on the merger
of binary neutron star systems terms have been observed in numerical
simulations of late
inspiral~\cite{Read2009a,Baiotti2010,Baiotti2011,Bernuzzi2012,Hotokezaka2013}.  Additional
information is also present in the frequencies of neutron-star normal modes
after the merger, should the EOS be stiff enough to support a hypermassive
neutron star~\cite{Shibata2005a,Bauswein2010a,Stergioulas2011}. In this paper,
we incorporate a wider range of EOS than previous work, with systematic
parameter variation that allows us to explicitly estimate how EOS parameters
will be constrained; we show that, to within the accuracy of our simulations,
the parameter $\Lambda$ can also be used to characterize the merger of binary
neutron stars.

While numerical-relativity efforts can simulate binary coalescence
during the highly dynamical phase at the endpoint of binary inspiral, there 
are additional challenges in determining the slow inspiral motion: simulating the
length-scales and time-scales of widely separated binary systems whose orbital
decay occurs over many cycles is computationally expensive, and the resolution
must be sufficiently high that the numerical scheme conserves angular momentum
and energy with enough accuracy that the relatively small gravitational
radiation dominates.  However, if high-quality numerical simulations extend
into the region in which a given post-Newtonian or other analytic approximation
is valid, joining the post-Newtonian waveform to the numerical waveform at a
point when both waveforms are deemed accurate will yield a complete {\it
hybrid} waveform for the binary system, which includes both inspiral tidal
effects and other hydrodynamic effects that occur during coalescence.
Hybrid waveforms can be used to better measure the EOS-dependent properties of
a neutron star. The measurement will be limited not only by statistical
errors arising from the fact that the gravitational-wave signal must be
extracted from detector noise, but also by systematic errors arising from
modeling errors in the analytic inspiral, the numerical simulations, and ambiguities in the process of joining them together. 

We also explore an alternative scenario:  If systematic errors
arising from hybrid waveforms are intractable, it is possible to use only
numerical simulations of the late inspiral, which \emph{are} robust, to
estimate structural parameters of the neutron star.  The measurability of the
tidal deformability suffers in such an approach because the unknown time and
phase of the numerical waveform relative to the time and phase of
post-Newtonian models of the early inspiral must be marginalized over.
However, we show that we are still able to constrain the tidal deformability of
neutron stars in binary neutron-star systems, using current numerical
simulations, even if hybrid waveforms cannot be constructed.  

Our results are derived from data produced by two
independent numerical-relativity codes: \sacra~\cite{Yamamoto2008} and 
\whisky~\cite{Baiotti03a,Baiotti04,Giacomazzo:2007ti}. This has the advantages of checking
the actual numerical differences due to different implementations of the equations (Einstein
equations, relativistic-hydrodynamics equations) and of understanding if and how much such
differences are relevant to gravitational-wave detection and analysis.

We use a spacelike signature $(-, +, +, +)$ and a system of units in which c = G = 1.   Greek indices are taken to run from 0 to 3, Latin indices from 1 to 3, and we adopt the standard convention for the summation over repeated indices.

\section{Generation of Waveforms}
\subsection{EOS variation}
\label{sec:eos}

We specify EOS candidates in the framework of~\cite{Read2009,Read2009a}: A fixed crust
EOS is joined to a core EOS that
we vary using a piecewise polytrope scheme. Currently we consider a single
core region, but we vary independently two parameters: the adiabatic index
$\Gamma$ of the core and the overall pressure scale $p_*$ at a fiducial
density $\rho_*=10^{14.7}$g/cm$^3$.   Following the notation of~\cite{Read2009a,Kyutoku2010}, we
categorize the EOS by the pressure scale: From high pressure to low
pressure we use 2H, H, HB, and B.
The adiabatic index variation is
indicated by one or more occurrences of a lower case {\em s}: H (no {\em s}) has $\Gamma=3$, 
Hs has $\Gamma = 2.7$, Hss has $\Gamma = 2.4$.  Eight EOSs (2H,H,
HB, B, Bs, Bss, HBs, HBss) were simulated using both the \whisky and \sacra
codes, at multiple resolutions and at different initial stellar separations
(see Sec.~\ref{sec:initial_data}). EOS parameters are summarized in Table \ref{tab:eos}.

\begin{table}
\caption{EOS parameters and properties of individual neutron
stars for the reference mass $1.35 M_\odot$ simulated in this
work. The parameter $p_*$ is measured in $\text{dyn}/\text{cm}^2$, $R$ is
measured in km, $C$ is the compactness ($M_\text{NS}/R$), and the tidal parameter $\lambda$ used in previous
work \cite{Hinderer2010} has units of $10^{36}$\,g$/$cm$^2$. \label{tab:eos}}
\begin{ruledtabular}
\begin{tabular}{lccccccc}
EOS 
& $\log_{10} p_*$
& $\Gamma$  & $R$ & $C$ &
$\lambda$
& $\Lambda^{1/5}$ &
$\Lambda$  \\ \hline
 2H & 34.9036 & 3.0 & 15.23& 0.131  & 10.97\phantom{00} & 4.713 & 2325\phantom{.0} \\
 H & 34.5036 & 3.0 & 12.28 & 0.162 & \phantom{0}2.866\phantom{0} & 3.603 & \phantom{0}607.3 \\
 HB & 34.4036 & 3.0 & 11.61 & 0.172 & \phantom{0}1.992\phantom{0} & 3.350 & \phantom{0}422.0 \\
 B & 34.3036 & 3.0 & 10.96 & 0.182 & \phantom{0}1.362\phantom{0}  & 3.105 & \phantom{0}288.7 \\
 Bs & 34.3036 & 2.7 & 10.74& 0.186 & \phantom{0}1.075\phantom{0}  & 2.961 & \phantom{0}227.7 \\
 Bss & 34.3036 & 2.4 & 10.27& 0.194 & \phantom{0}0.6695 & 2.694 & \phantom{0}141.9 \\
 HBs & 34.4036 & 2.7 & 11.58& 0.172 & \phantom{0}1.770\phantom{0}  & 3.275 & \phantom{0}376.9 \\
 HBss & 34.4036 & 2.4 & 11.45& 0.174  & \phantom{0}1.421\phantom{0}  & 3.131 & \phantom{0}301.1 \\
\end{tabular}
\end{ruledtabular}
\end{table}
\begin{figure}[tb]
\includegraphics[width=8.5cm]{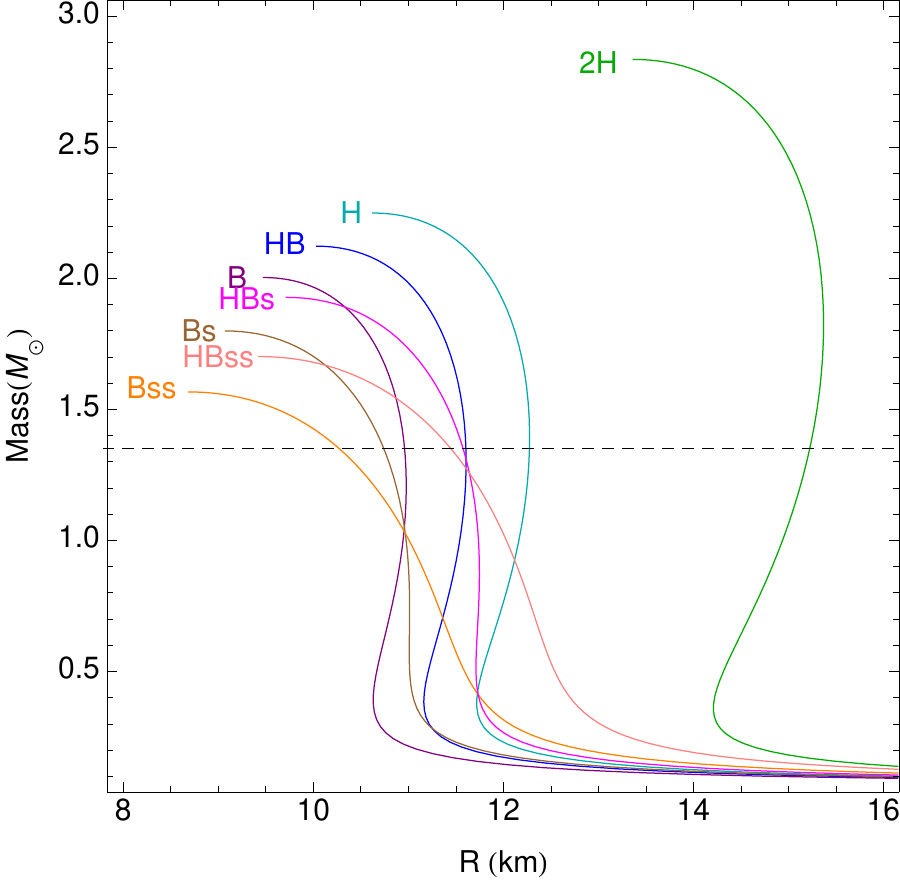}
\caption{The radius $R$ of the simulated EOS as a function of mass. The dashed lines indicated the simulated mass value of 1.35\,$M_\odot$.}
\label{fig:radlambda}
\end{figure}

For a given neutron-star mass, each EOS can be identified by two useful
macroscopic characteristic quantities, $R$ and $\Lambda$: $R$ is the stellar
radius of an isolated nonrotating neutron star and 
\begin{equation}
\Lambda \equiv \frac{2}{3} k_2 \left(\frac{R}{M}\right)^5
\label{eq:Lambda}\end{equation}
is the dimensionless quadrupole tidal deformability ($k_2$ is the quadrupole
Love number). These parameters are tabled for the current models in
Table~\ref{tab:eos}.  Recent analysis of neutron-star matter properties
compatible with modern nuclear theory~\cite{Hebeler2010} suggests that the
radius of the 2H model is unrealistically large, and the neutron-star mass
measurement of 2.0 $M_\odot$\cite{Demorest2010} rules out the ``s'' EOSs.
Current astrophysical
constraints~\cite{Steiner2010,guver11,ozel09,ozel10a,ozel10b,suleimanov10}
further favor EOS H and HB. However, we consider this range useful for a
parameter study.

At leading order in the 
separation of the stars, $\Lambda$ determines the $(\ell,m)=(2,0)$ departure 
of the asymptotic metric from spherical symmetry
and the departure of the waveform phase evolution from its point-particle form.
Our results imply that $\Lambda$ effectively determines the waveform's
departure from point-particle (or nonspinning BH-BH) inspiral even for the
late inspiral.

\begin{figure}[tb]
\includegraphics[width=8.5cm]{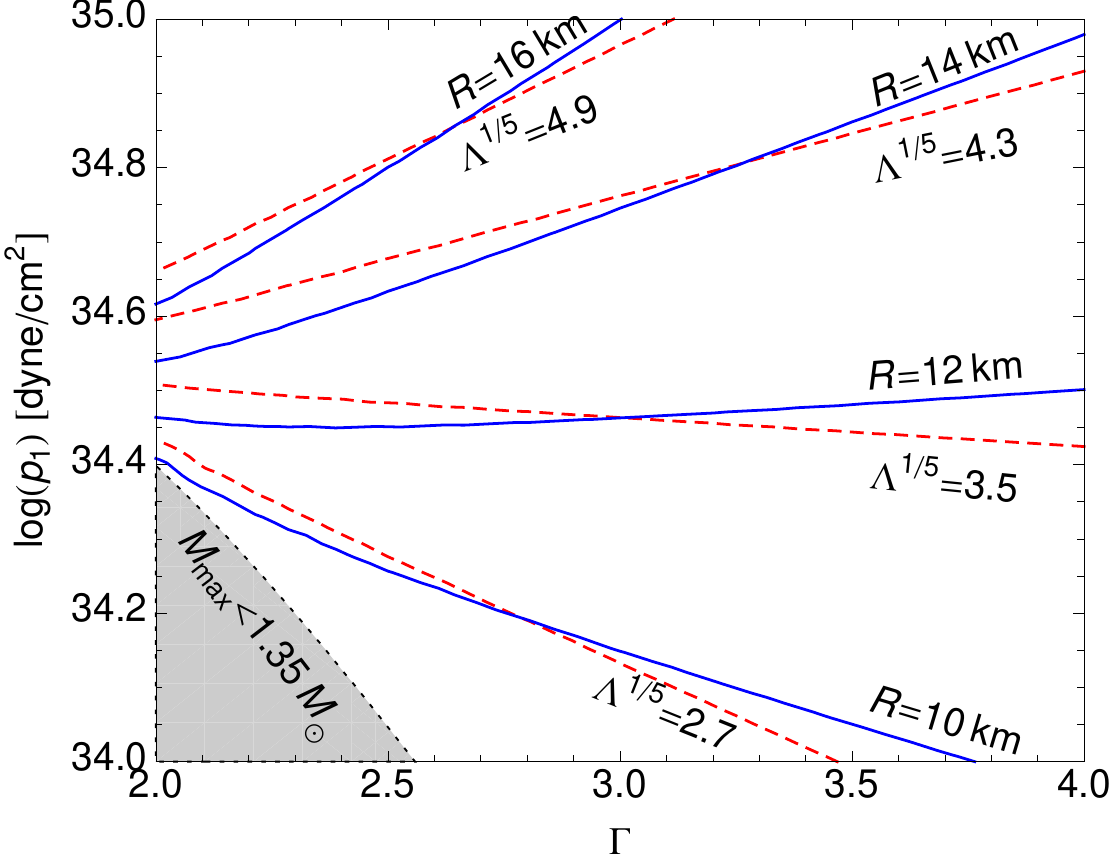}
\caption{Contours of constant $R$ and $\Lambda$ (labeled by 
the value of $\Lambda^{1/5}$) in the two-parameter EOS space.}
\label{fig:eos-Lambda-R}
\end{figure}

Fig.~\ref{fig:eos-Lambda-R} (provided by B.~D.~Lackey) shows contours of 
constant $R$ and $\Lambda$ for 1.35\,$M_\odot$ stars in the EOS space.  Our
simulations suggest that the contours in the EOS parameter space of constant
departure of the waveform from point-particle inspiral coincide with similar
accuracy with these contours of constant $\Lambda$, but the range of
high-resolution runs is not yet large enough for a quantitative conclusion.  

\subsection{Numerical simulations}\label{ss:numericalsimulations}

Here we give only a brief overview of the codes, while we refer the reader to
previous articles for more
details~\cite{Taniguchi10,Yamamoto2008,Baiotti03a,Baiotti04,Pollney:2007ss,Baiotti08}. 

\subsubsection{Initial data}
\label{sec:initial_data}

The initial configurations for our simulations are produced using the numerical
code of~\cite{Gourgoulhon01,Taniguchi02b,Taniguchi03}, based on the multidomain
spectral-method library, \programname{LORENE}\@. \programname{LORENE} was
originally written by the Meudon relativity group~ and is publicly
available~\cite{lorene}.  We have added a new method to treat the piecewise
polytropic EOS of Sec.~\ref{sec:eos}, which is used in \cite{Taniguchi10} for
detailed study of quasiequilibrium sequences with such EOS.

The total mass is fixed to be $M = M_{\rm tot} = 2.7 M_{\odot}$ at infinite
separation. We consider equal-mass binaries. The initial data are prepared for
two different orbital angular velocities, $M \Omega_0 = 0.0188$ and
0.0221, where $\Omega_0$ denotes the initial orbital angular velocity, subsequently labelled by ``I188'' and ``I221''.  Nine
models are prepared for our simulations, varying the EOS and orbital angular
velocity for fixed total mass. Some of the physical quantities of the initial
configurations are reported in Table~\ref{table:ID}.

\begin{table*}[tb]
  \caption{
    Properties of the initial data: proper separation between
    the centers of the stars $d/\tilde{M}_{_{\rm ADM}}$; baryon mass $M_{b}$
    of each star in units of solar mass; total ADM mass $M_{_{\rm ADM}}$ in
    units of solar mass, as measured on the finite-difference grid with the \whisky code and with
    the \sacra code; total ADM
    mass $\tilde{M}_{_{\rm ADM}}$ in units of solar mass, as provided by the
    Meudon initial data; angular momentum $J$, as measured on the
    finite-difference grid with the \whisky code and with
    the \sacra code; angular momentum $\tilde{J}$, as provided
    by the Meudon initial data; initial orbital angular velocity expressed as 
    $\tilde{M}_{\rm ADM}\Omega_0$; mean coordinate equatorial radius of each star $r_e$
    along the line connecting the two stars; maximum rest-mass density
    of a star $\rho_{\rm max}$. Note that the
    values of $M_{\rm ADM}$ and $J$ are computed through a volume
    integral in \whisky, while in \sacra they are computed through the extrapolation to $r \rightarrow
    \infty$ of the ADM masses and angular momenta calculated as surface integrals at finite radii $r$.}
\begin{ruledtabular}
\catcode`?=\active
\def?{\kern\digitwidth}
\begin{tabular}{lccccccccccccc}
	& &
	\multicolumn{4}{c}{Mass ($M_\odot$)} &
	\multicolumn{3}{c}{$J$ $(\times10^{49}{\rm g\, cm}^2{\rm /s})$} &
 &
	$r_e$ &
	$\rho_{\rm max}$ \\
\cline{3-6} \cline{7-9}
EOS & $d/\tilde{M}_{_{\rm ADM}}$ & $M_b$ &
	$M_{\text{ADM}}^{\smash{\whisky}}$ &
	$M_{\text{ADM}}^{\smash{\sacra}}$ &
	$\tilde{M}_{\text{ADM}}$ &
	\whisky & \sacra & $\tilde{J}$ &
	$\tilde{M}_{_{\rm ADM}}\Omega_0$ &
	$({\rm km})$ &
	$({\rm g/cm}^3)$ \\
\hline
2H I188 & $13.4$  & $1.455$  & $2.671$ & $2.682$ & $2.678$  
& $6.772$ & $6.781$ & 
$6.772$  & $0.0187$  & $12.99?$  & $3.74\times10^{14}$\\
HB I188 & $13.5$  & $1.493$  & $2.671$ & $2.682$ & $2.678$  
& $6.761$ & $6.769$ & 
$6.761$  & $0.0186$  & $?9.218$  & $8.27\times10^{14}$\\
B I221& $15.4$  & $1.502$  & $2.668$ & $2.680$ & $2.675$  
& $6.492$ & $6.499$ & 
$6.491$  & $0.0219$  & $?8.48?$  & $9.77\times10^{14}$\\
Bss I221 & $11.9$  & $1.501$  & $2.669$ & $2.680$ & $2.675$  
& $6.493$ & $6.501$ & 
$6.493$  & $0.0219$  & $?7.85?$  & $1.49\times10^{15}$\\
\end{tabular}
\end{ruledtabular}
\vskip -0.25cm
\label{table:ID}
\end{table*}

\subsubsection{Overview of evolution codes}

Both the \sacra and \whisky codes evolve the Einstein equations in the
Baumgarte-Shapiro-Shibata-Nakamura formalism~\cite{Nakamura87,
Shibata95, Baumgarte99, Alcubierre99d}. For the \whisky simulations, 
the Einstein equations are solved using the \programname{CCATIE} code, a three-dimensional finite-differencing code
based on the \programname{Cactus Computational Toolkit}~\cite{Goodale02a}. A
detailed presentation of the \programname{CCATIE} code and of its convergence properties
has been presented in~\cite{Pollney:2007ss}. For tests and details
on \sacra, see~\cite{Yamamoto2008}.

The gauges are specified in terms of the standard Arnowitt-Deser-Misner (ADM) lapse
function, $\alpha$, and shift vector, $\beta^i$~\cite{misner73}.
We evolve the lapse according to the ``$1+\log$'' slicing
condition~\cite{Bona94b}:
\begin{equation}
  \partial_t \alpha - \beta^i\partial_i\alpha 
    = -2 \alpha K.
  \label{eq:one_plus_log}
\end{equation}
The shift is evolved using the hyperbolic $\tilde{\Gamma}$-driver
condition~\cite{Alcubierre02a}
\begin{eqnarray}
\label{shift_evol}
  \partial_t \beta^i - \beta^j \partial_j  \beta^i & = & \frac{3}{4} B^i\,,
  \\
  \partial_t B^i - \beta^j \partial_j B^i & = & \partial_t \tilde\Gamma^i 
    - \beta^j \partial_j \tilde\Gamma^i - \eta B^i\,,
\end{eqnarray}
where $B^i$ is an auxiliary variable and $\eta$ is a parameter that acts as a
damping coefficient. We set
$\eta=1.0$ or $\eta\approx 0.5$, in units of $M_\odot=1$, for \whisky and
\sacra respectively.

Both codes 
adopt a \textit{flux-conservative} formulation of the hydrodynamics
equations~\cite{Marti91,Banyuls97,Ibanez01}, in which the set of
conservation equations for the stress-energy tensor $T^{\mu\nu}=\rho h u^\mu u^\nu+p g^{\mu\nu}$ and
for the matter current density $J^\mu=\rho u^{\mu}$ (where $p$ is the pressure, $\rho$ is the rest-mass
density, $\varepsilon$ is the specific internal energy, {$h \equiv 1 + \varepsilon + p/\rho$} is the specific enthalpy, 
$u^\mu$ is the four-velocity and $g^{\mu\nu}$ is the inverse metric), namely
%
$\nabla_\mu T^{\mu\nu} = 0$ and 
$\nabla_\mu J^\mu = 0$,
are written in a hyperbolic, first-order, flux-conservative form of
the type
\begin{equation}
\label{eq:consform1}
\partial_t {\mathbf q} + 
        \partial_i {\mathbf f}^{(i)} ({\mathbf q}) = 
        {\mathbf s} ({\mathbf q})\ ,
\end{equation}
where ${\mathbf f}^{(i)} ({\mathbf q})$ and ${\mathbf s}({\mathbf q})$
are the flux vectors and source terms, respectively~\cite{Font03}.
The EOS closes the system by relating
pressure, rest-mass density, and internal-energy density.

The system written in conservative form is solved with high-resolution shock-capturing methods, in
several variants for both codes. For the simulations of this work, both codes employ 3rd order
piecewise-parabolic method (PPM)~\cite{Colella84} reconstruction, but \sacra used Kurganov-Tadmor's central scheme~\cite{Tadmor00}
for Riemann solvers, while \whisky~\cite{Aloy99b} use the Marquina flux formula.
The details of the differences in the implementations of the Einstein 
and hydrodynamics equations in the two codes are described in~\cite{Baiotti2010a}, which contains
also convergence tests and the description of the differences in the implementations of 
adaptive mesh refinement~\cite{Schnetter-etal-03b,Yamamoto2008}.

\begin{table*}[tb]
  \caption{ Properties of the initial grids: the model name has the format: EOS name - code used for
    simulation - resolution (R\%, where \% is the spacing of the finest grid in meters) -
    initial frequency (I\%); $n$ is the number of refinement levels (including the coarsest grid);
    $m$ is the number of finer levels that are moved to follow the stars; $h_{\text{fine}}$ is the
    spacing of the finest level; $L_{\text{fine}}$ is the length of the side of the finest level;
    $h_{\text{course}}$ is the spacing of the coarsest level; $r$ is the outer-boundary
    location. All lengths are expressed in km. 
}
\begin{ruledtabular}
\begin{tabular}{lcccccc}
       & Refinement & Moving & \multicolumn{2}{c}{Finest Grid (km)} &
  \multicolumn{2}{c}{Coarsest Grid (km)} \\
\cline{4-5} \cline{6-7}
Model  & Levels $n$ & Levels $m$ & Spacing $h_{\text{fine}}$ & Extent $L_{\text{fine}}$ &
 Spacing $h_{\text{coarse}}$ & Outer Boundary $r$ \\
\hline
B \whisky R141 I221 (HR) & $6$ & 2 & $0.1418$ & $44.33$ & $\ \ 4.54$ & $760$ \\
B \whisky R177 I221 (MR) & $6$ & 2 & $0.1773$ & $44.33$ & $\ \ 5.67$ & $760$ \\
B \whisky R221 I221 (LR) & $6$ & 2 & $0.2216$ & $44.33$ & $\ \ 7.09$ & $760$ \\
B \sacra  R157 I221 (HR) & $7$ & 4 & $0.1570$ & $\ \ \ \ 9.420$ & $10.05$ & $603$ \\
B \sacra  R174 I221 (MR) & $7$ & 4 & $0.1744$ & $\ \ \ \ 9.420$ & $11.16$ & $603$ \\
B \sacra  R202 I221 (LR) & $7$ & 4 & $0.2023$ & $10.12$ & $12.95$ & $648$ \\
\end{tabular}
\end{ruledtabular}
\vskip -0.25cm
\label{table:grids}
\end{table*}


For the highest-resolution runs with \whisky, the spacing of the
finest of the six grid levels is $h_{\text{fine}}=0.096\,M_{\odot}\approx 0.1418\,\km$
and the spacing in the wave zone (the coarsest grid) is
$h_{\text{coarse}}=3.072\,M_{\odot}\approx 4.536\,\km$. The finest grid always covers 
the whole stars. The outer boundary is located at about $760\ \km$.

For the runs with \sacra, the computational domain comprises
seven grid levels, with finest grid resolution 
$h_{\text{fine}}=0.1063\,M_{\odot} \approx 0.1570\,\km$ and with spacing in the wave
zone (the coarsest grid) $h_{\text{coarse}} = 6.804\,M_{\odot}\approx 10.05$ km for the 
highest-resolution runs. The finest grid covers the stellar radius completely 
(the boundary of the finest
grid is at $\approx 115\%$ of the stellar radius). The radius of the outer
boundary is about $603\ \km $.

The properties of the grids adopted in the simulations with the two codes are
summarized in Table~\ref{table:grids}. In general, we use a naming convention
to label results for a given numerical simulation, e.g.~``HB \whisky R141
I221'',  which summarizes the EOS (HB), the code (\whisky), the
resolution of the finest grid in meters (141), and the initial orbital angular
velocity imposed for building the initial data expressed as $\tilde{M}_{\rm ADM}\Omega_0*10^4$ (221).

\subsection{Waveform extraction}\label{ss:waveformextraction}

This work is concerned primarily with the gravitational waveform extracted from
the simulation, rather than the underlying density or pressure distributions,
so we describe in some detail the gravitational-wave methods employed.

Both codes compute the gravitational 
waveforms using the Newman-Penrose formalism~\cite{Newman62a}, which provides a
convenient representation for a number of radiation-related quantities
as spin-weighted scalars. In particular, the curvature scalar
\begin{equation}
  \Psi_4 \equiv -C_{\alpha\beta\gamma\delta}
    n^\alpha \bar{m}^\beta n^\gamma \bar{m}^\delta
  \label{eq:psi4def}
\end{equation}
is defined as a particular component of the Weyl curvature tensor
$C_{\alpha\beta\gamma\delta}$ projected onto a given null frame
$\{\boldsymbol{l}, \boldsymbol{n}, \boldsymbol{m},
\bar{\boldsymbol{m}}\}$ and can be identified with the gravitational
radiation field if a suitable frame is chosen at the extraction radius. In
practice, we define an orthonormal basis in the three-space
$(\hat{\boldsymbol{r}}, \hat{\boldsymbol{\theta}},
\hat{\boldsymbol{\phi}})$, centered on the Cartesian origin and
oriented with poles along $\hat{\boldsymbol{z}}$. The normal to the
slice defines a timelike vector $\hat{\boldsymbol{t}}$, from which we
construct the null frame
\begin{equation}
   \boldsymbol{l} = \frac{1}{\sqrt{2}}(\hat{\boldsymbol{t}} - \hat{\boldsymbol{r}}),\quad
   \boldsymbol{n} = \frac{1}{\sqrt{2}}(\hat{\boldsymbol{t}} + \hat{\boldsymbol{r}}),\quad
   \boldsymbol{m} = \frac{1}{\sqrt{2}}(\hat{\boldsymbol{\theta}} - 
     {\mathrm i}\hat{\boldsymbol{\phi}}) \ .
\end{equation}
We then calculate $\Psi_4$ via a reformulation of (\ref{eq:psi4def}) 
in terms of ADM variables on the slice~\cite{Shinkai94}:
\begin{equation}
  \Psi_4 = C_{ij} \bar{m}^i \bar{m}^j,  \label{eq:psi4_adm}
\end{equation}
where
\begin{equation}
  C_{ij} \equiv R_{ij} - K K_{ij} + K_i{}^k K_{kj} 
    - {\rm i}\epsilon_i{}^{kl} \nabla_l K_{jk}
\end{equation}
and $\epsilon_{ijk}$ is the Levi-Civita symbol.
The gravitational-wave polarization amplitudes $h_+$ and $h_\times$
are then related to $\Psi_4$ by time
integrals~\cite{Teukolsky73}:
\begin{equation}
\ddot{h}_+ - {\rm i}\ddot{h}_{\times}=\Psi_4 \ ,
\label{eq:psi4_h}
\end{equation}
where the double overdot stands for the second-order time derivative.
Care is needed when performing such time
integrals~\cite{Baiotti:2008nf,Baiotti2011,Reisswig:2011}. In \sacra, they are
computed with the fixed-frequency integration method~\cite{Kyutoku2011}.

For the extraction of the gravitational-wave signal, each code implements a
second independent method
that is based on expressions involving
the  gauge-invariant metric perturbations of a spherically symmetric background 
spacetime~\cite{Moncrief74}. The wave data obtained in this way give results compatible 
with those obtained with the Newman-Penrose formalism and are not reported here.

We use only the $(\ell,m)=(2,2)$ mode in this work. For the
equal-mass cases considered, other modes are much smaller. 
The waveform is analyzed as a function of retarded time
$t = t_{\text{sim}} - r - 2 M_0 \ln(r/M_0)$
where $M_0$ is the ADM mass of the system at the initial time of the simulation.

We will use the complex combination of the  
extracted polarizations 
\begin{equation}
h \equiv h_+ - {\rm i} h_\times = |h| {\rm e}^{{\rm i}\phi} 
\end{equation}
in further analysis.  Some relevant quantities of a detected signal can be
calculated for either polarization and in this paper we will always show
the average result for both polarizations. 

An instantaneous frequency is extracted by taking the time derivative of the
phase $\phi$ of the complex waveform. The total accumulated phase is
reconstructed  by integrating the instantaneous frequency in subsequent phase plots.

The physical system simulated is the same under translations by arbitrary
parameters, $t_0$ and $\phi_0$, which describe, respectively, the time of the
start of the simulation relative to some reference time and the initial phase
of the simulation relative to some reference phase.  When comparing two
waveforms, these free parameters amount to a relative time shift and phase
shift between the waveforms.  For the numerical waveforms with the same initial
separation, one can take the time and phase to be zero at the retarded time
corresponding to the start of the simulation. For simulations of EOS B,
detailed comparison results between \sacra and \whisky are presented
in~\cite{Baiotti2010a}.

\begin{figure*}[tb]
\begin{tabular}{cc}
\includegraphics[width=8.8cm, trim=1.7cm 0cm  1.7cm 0cm ,clip=true]{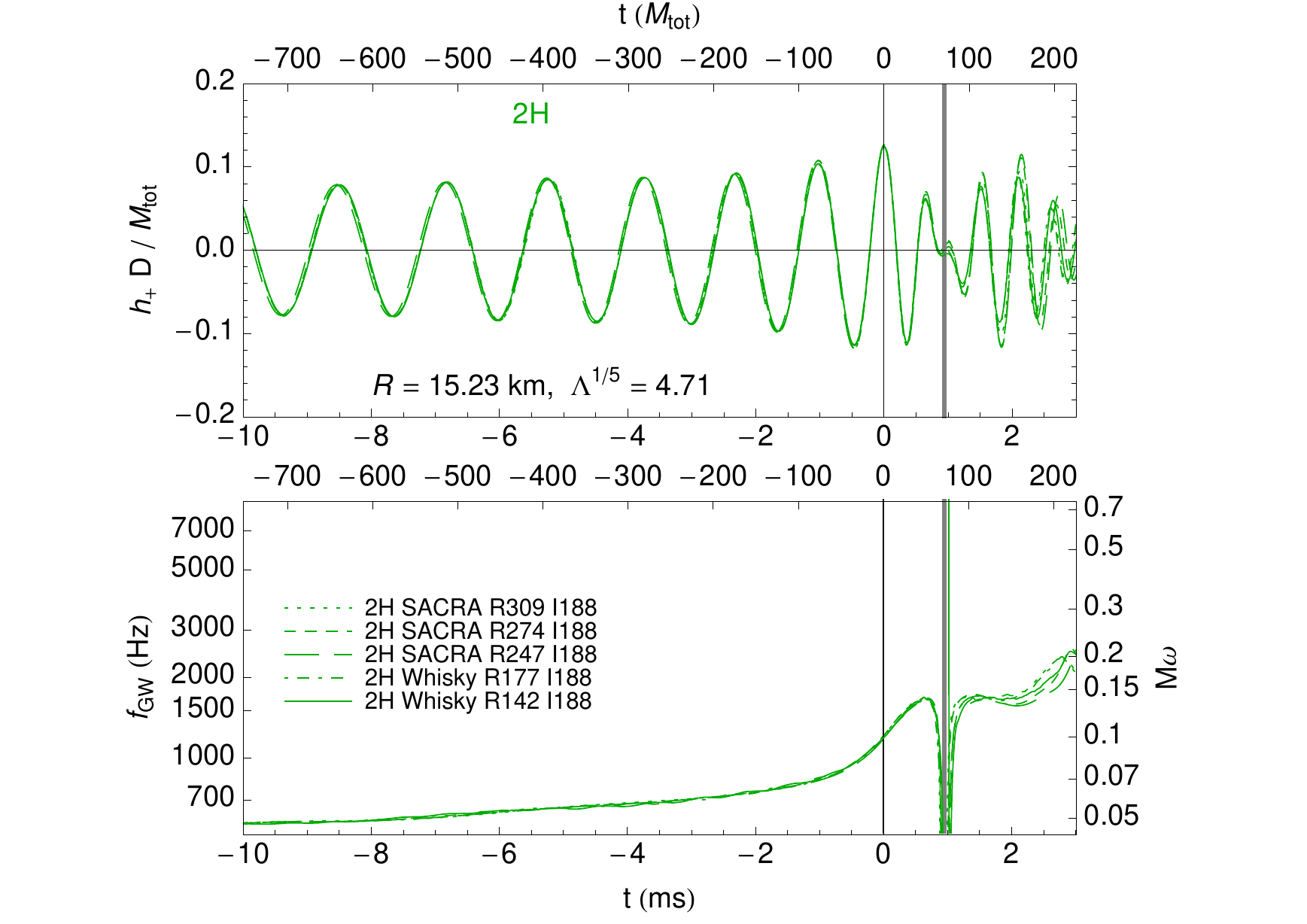} &
\includegraphics[width=8.8cm, trim=1.7cm 0cm 1.7cm 0cm ,clip=true]{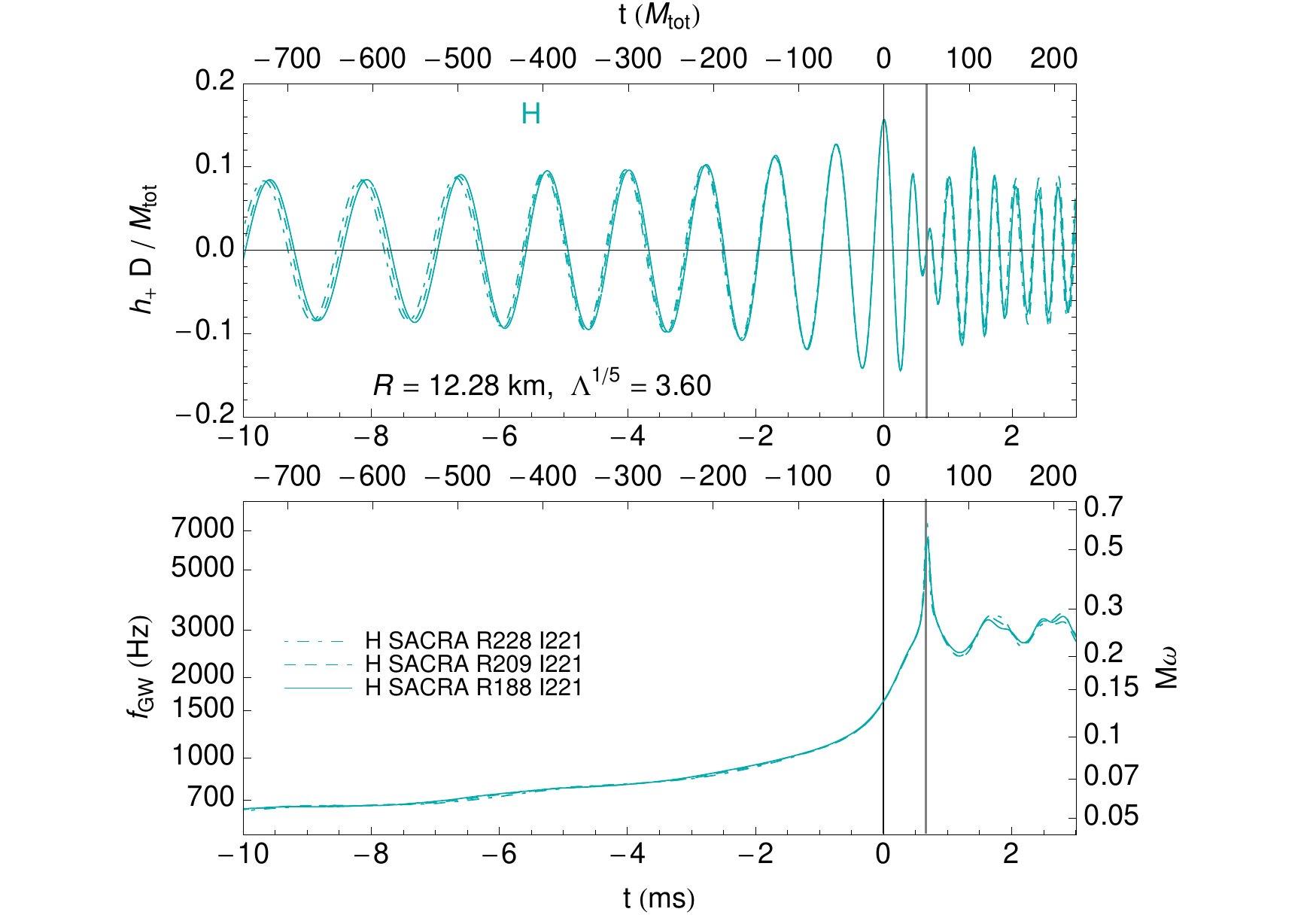} \\
\includegraphics[width=8.8cm, trim=1.7cm 0cm 1.7cm 0cm ,clip=true]{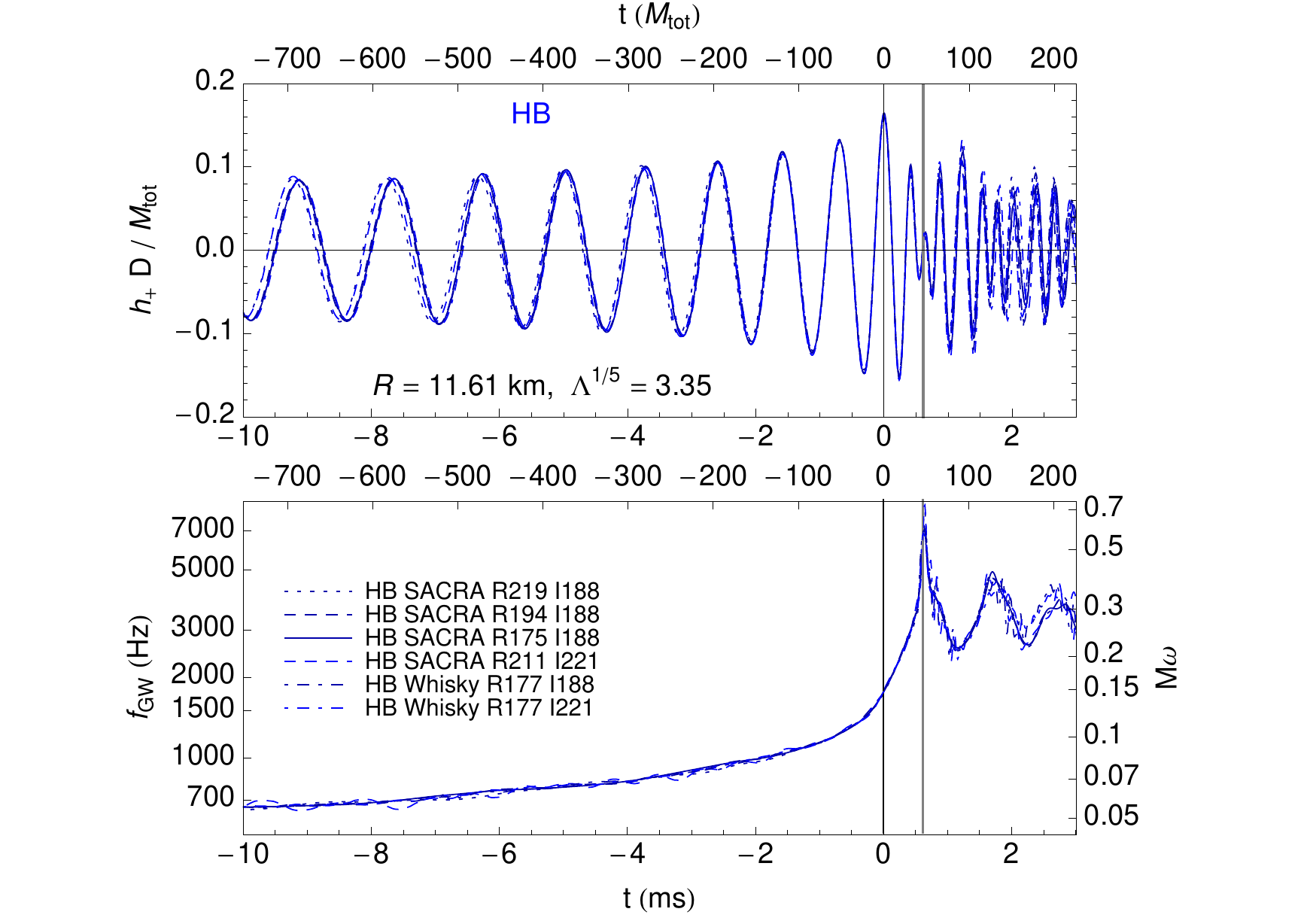} &
\includegraphics[width=8.8cm, trim=1.7cm 0cm 1.7cm 0cm ,clip=true]{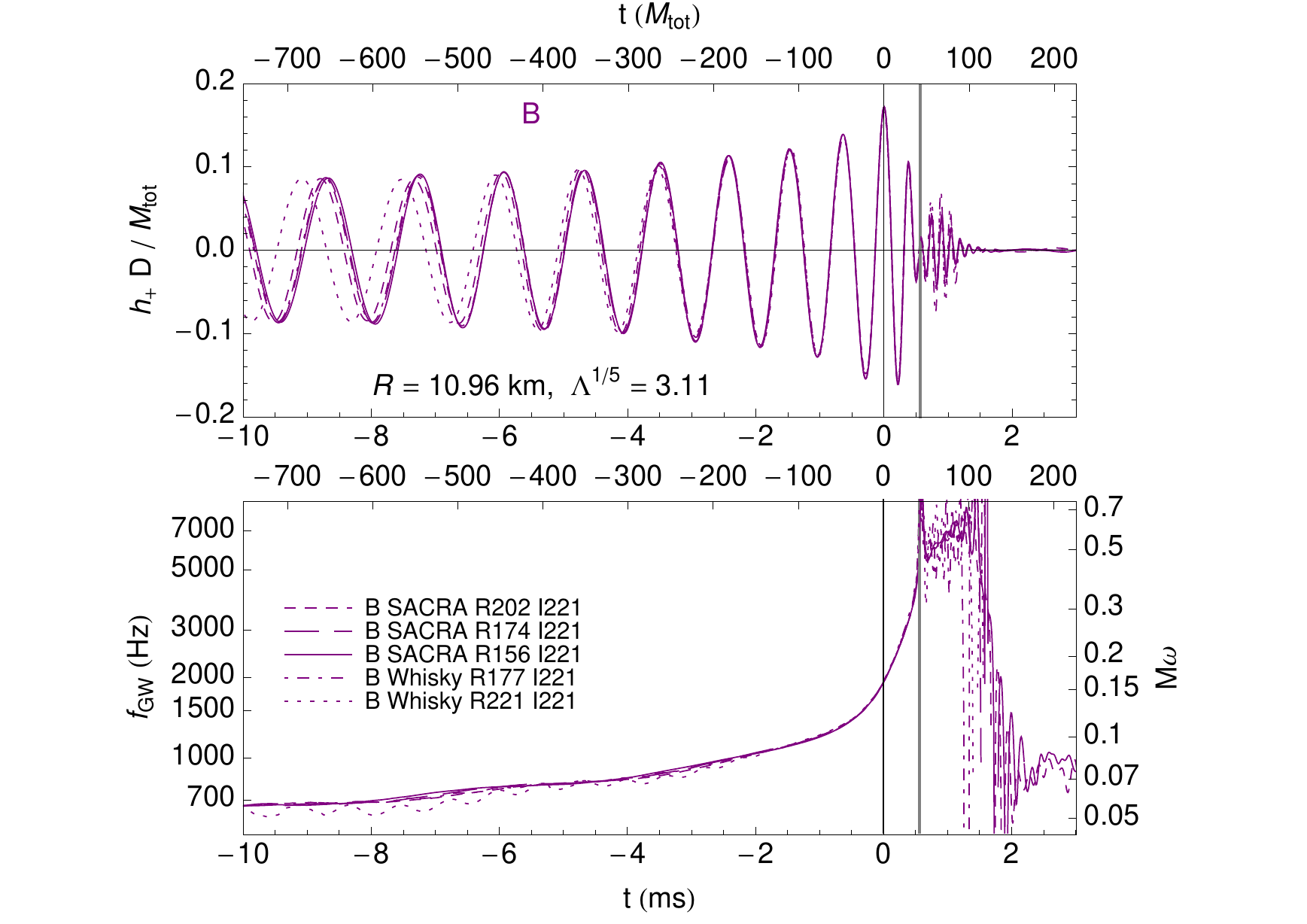} \\
\includegraphics[width=8.8cm, trim=1.7cm 0 1.7cm 0.1cm ,clip=true]{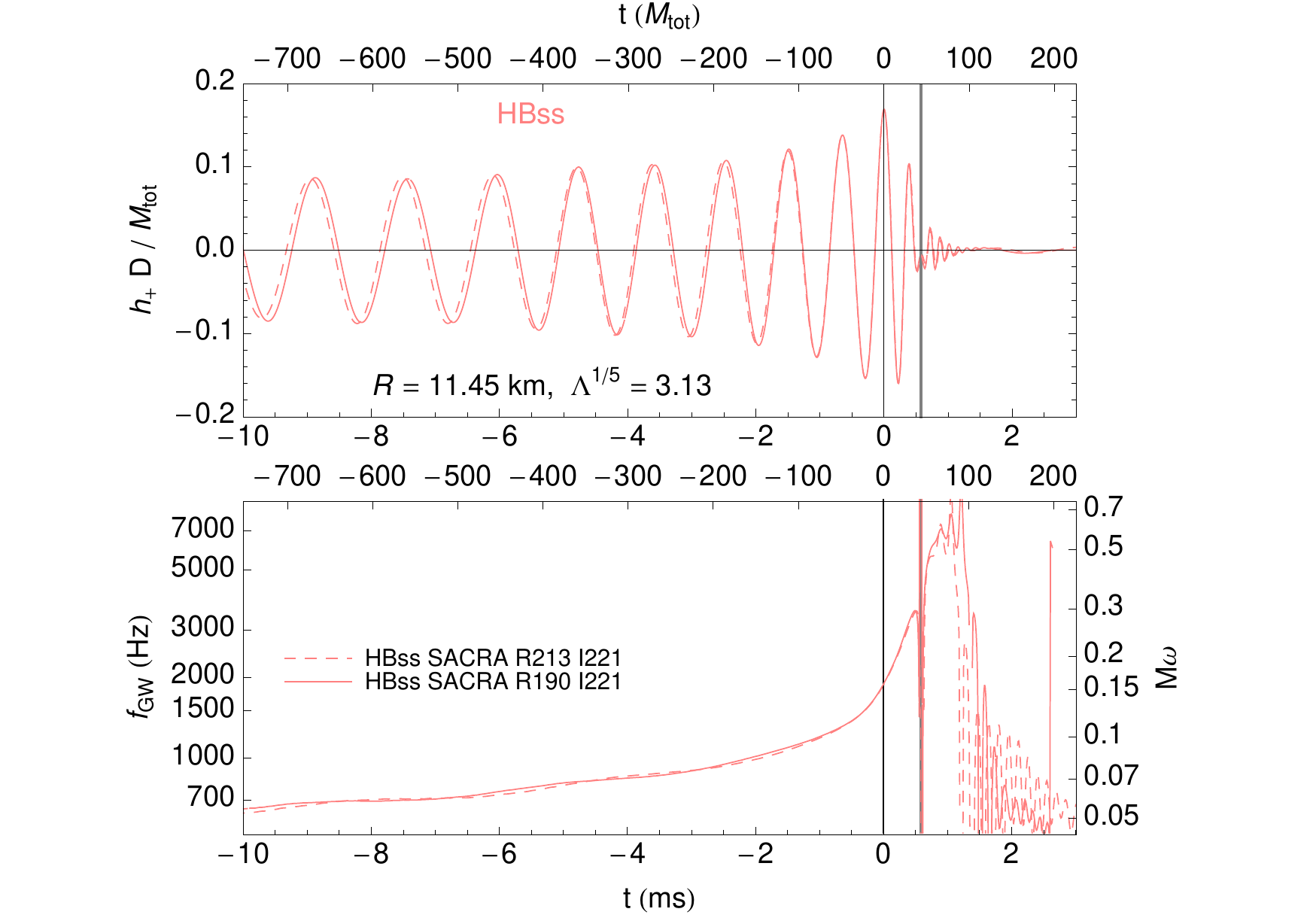} &
\includegraphics[width=8.8cm, trim=1.6cm 0 1.8cm 0 ,clip=true]{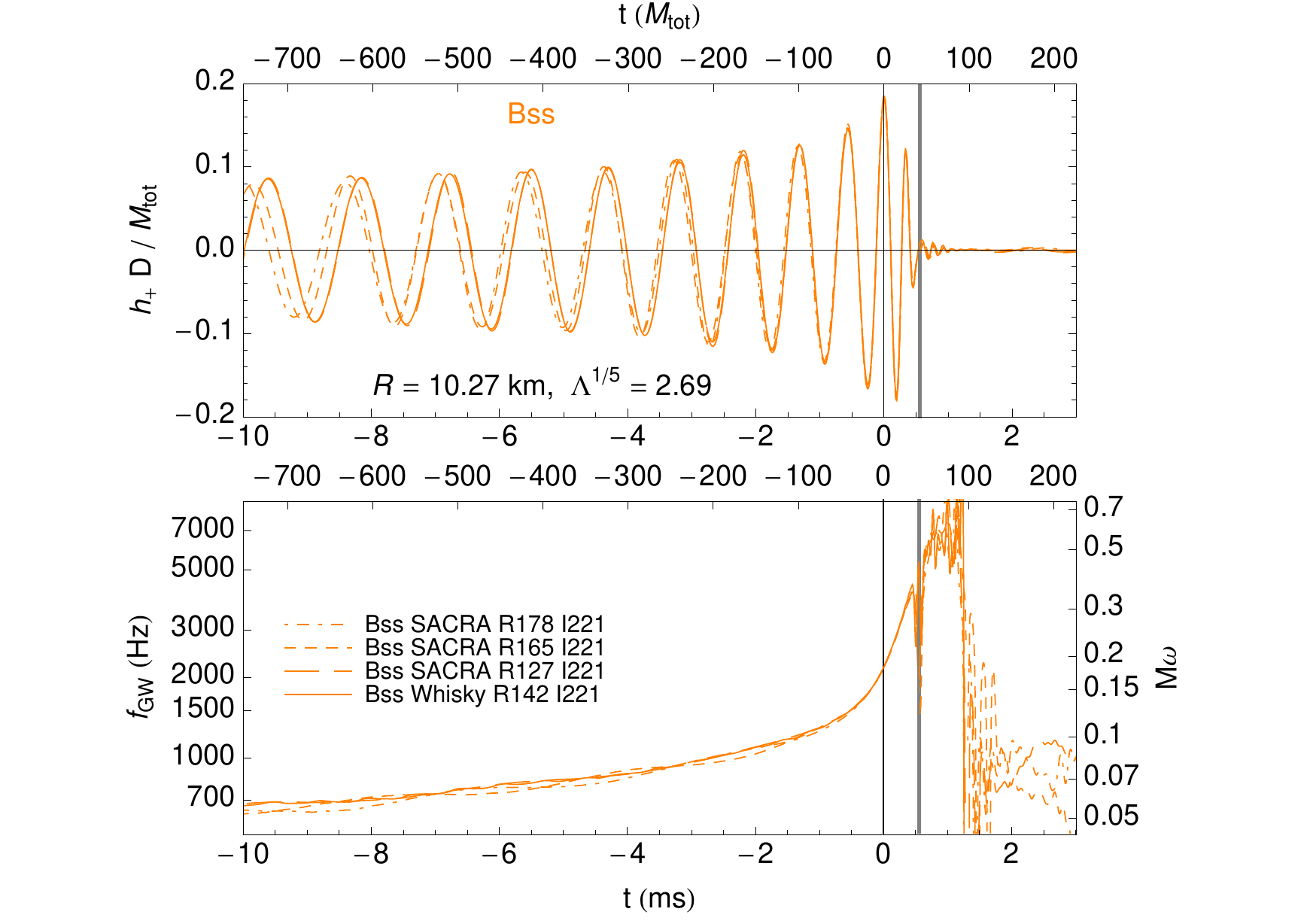} \\
\end{tabular}
\caption{Waveforms and time-frequency relations near merger, for the set of
simulations. We fix $t=0$ and $\phi = \arg h=0$ for all waveforms
at the peak amplitude point (see text). Time and phase are shown on the top
and bottom horizontal scales. Times of minimum amplitude are marked with
vertical grey lines, typically overlapping for simulations of the same
EOS\@. Instantaneous frequency is not well-defined in the neighborhood of minimum
amplitude (spikes or troughs there are spurious).
\label{fig:mergalign}}
\end{figure*}

\subsection{Common structure of waveforms}\label{ss:peak}

In order to compare simulations with different starting points, or to
remove artificial effects of initial data, an
alternate alignment procedure is required. 
In fully realistic binary simulations, as in the binary black hole case, the
merger is the simplest reference point for waveform comparison.  With finite
resolution, numerical dissipation may cause angular momentum to be artificially
lost during the evolution, increasing the rate of orbital decay during the
secular inspiral in a way that may mimic the tidal effects that are being
studied here.  Evolutions with different resolutions of the same initial data
tend to diverge from one another when they are aligned to start at the same
time and with the same phase.  However, this direct comparison 
overemphasizes differences that are less relevant to our purposes, as small
differences in phase accumulation in the early, low-frequency regime will
induce a corresponding time shift which translates to a large phase difference
in the later high-frequency cycles.  During the late stages of binary
coalescence, which are driven by dynamical effects, the effects of numerical
dissipation are less significant.  Comparable resolution-dependent features of
binary-black hole simulations motivated alignment of waveforms at merger in
comparisons such as those described in \citet{Hannam2009}.

\begin{figure*}[htb!]
\begin{tabular}{cc}
\includegraphics[width=8.5cm]{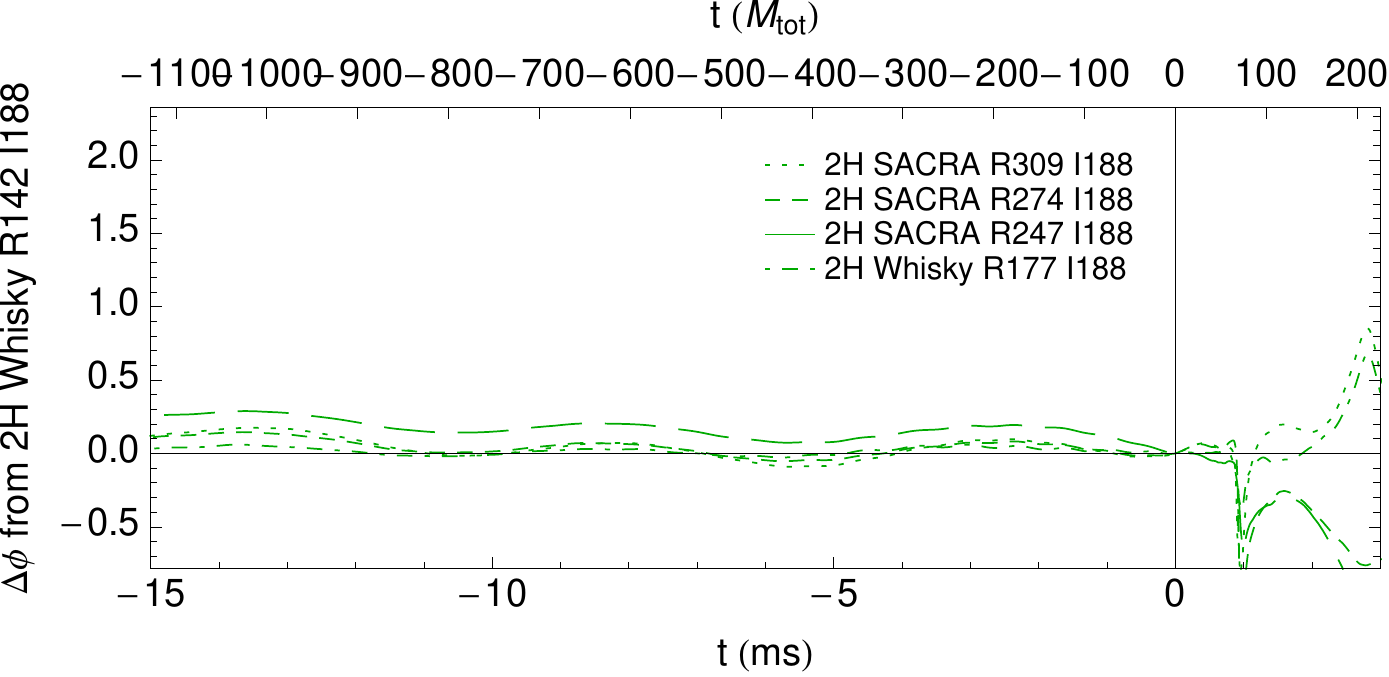} &
\includegraphics[width=8.5cm]{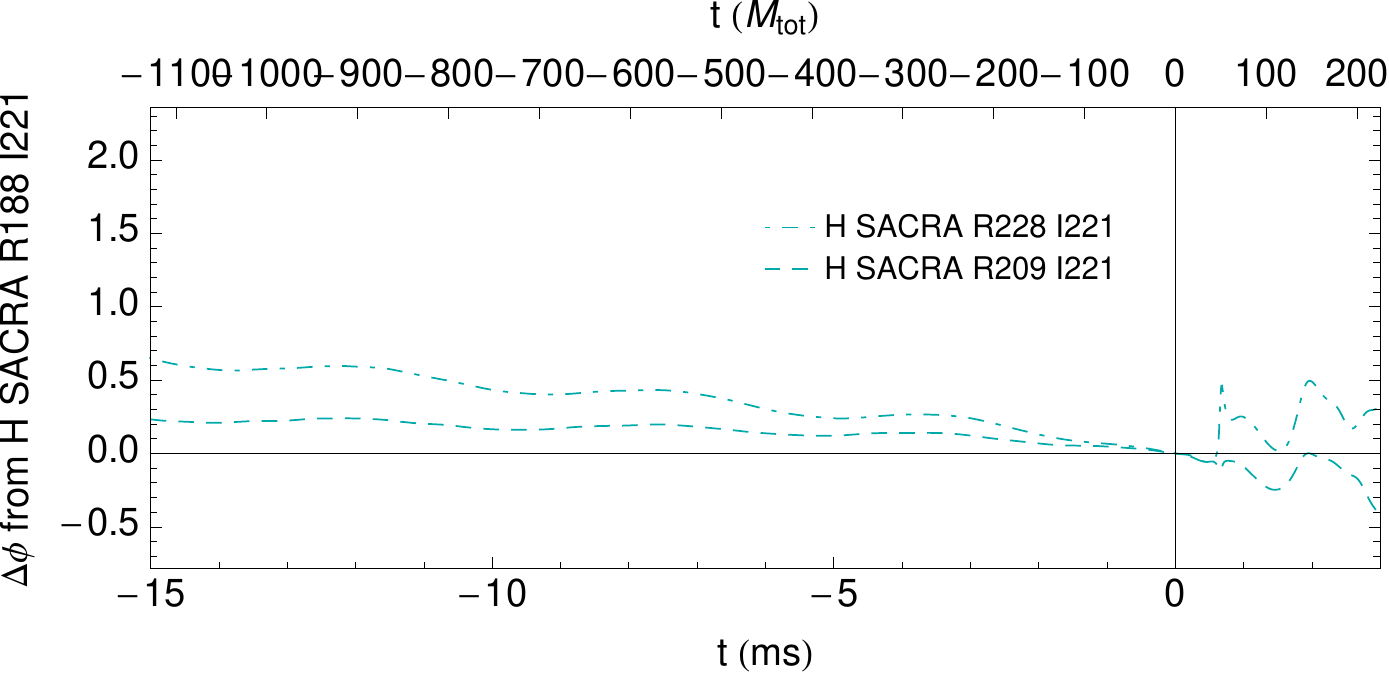} \\
\includegraphics[width=8.5cm]{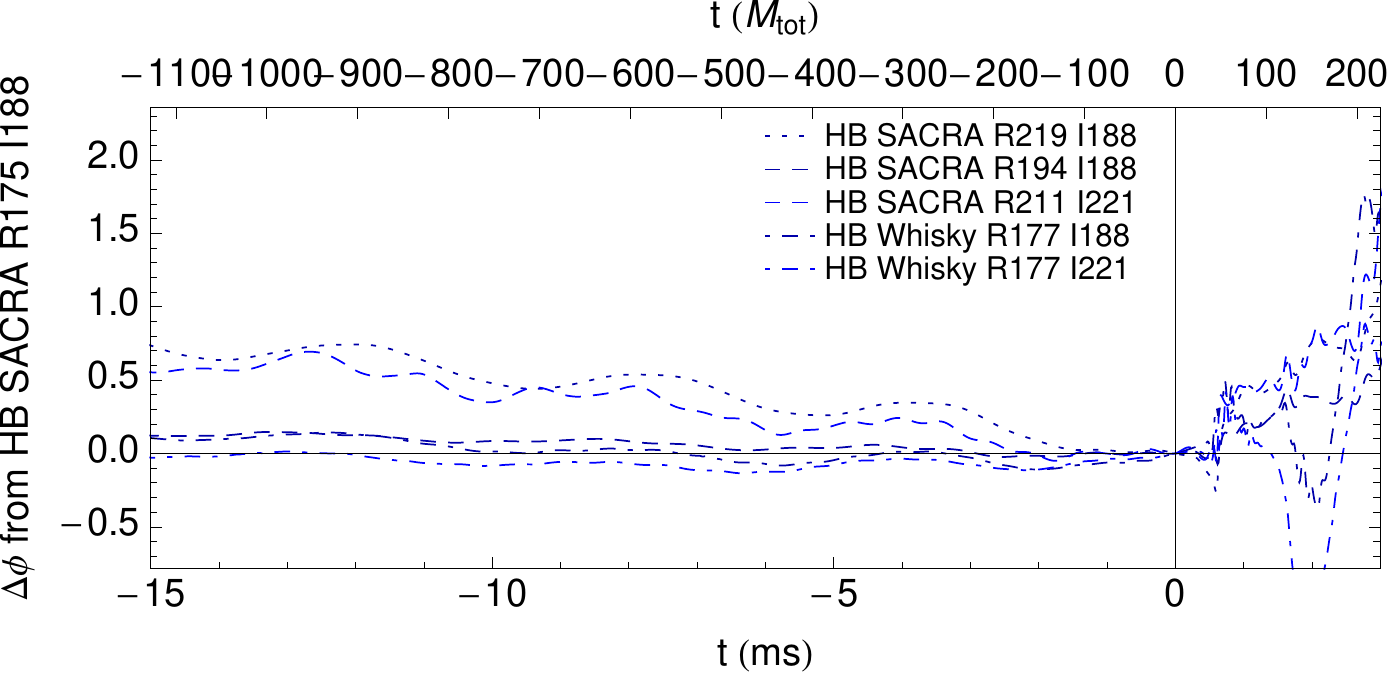} &
\includegraphics[width=8.5cm]{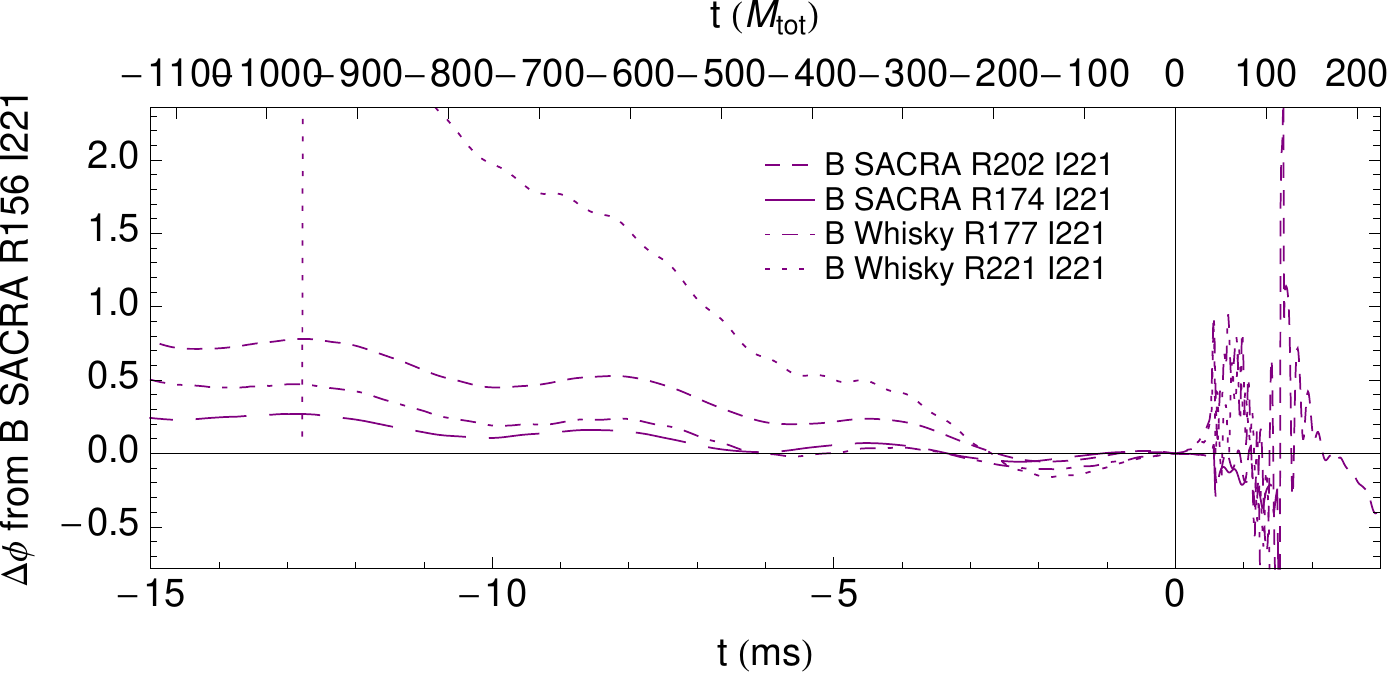} \\
\includegraphics[width=8.5cm]{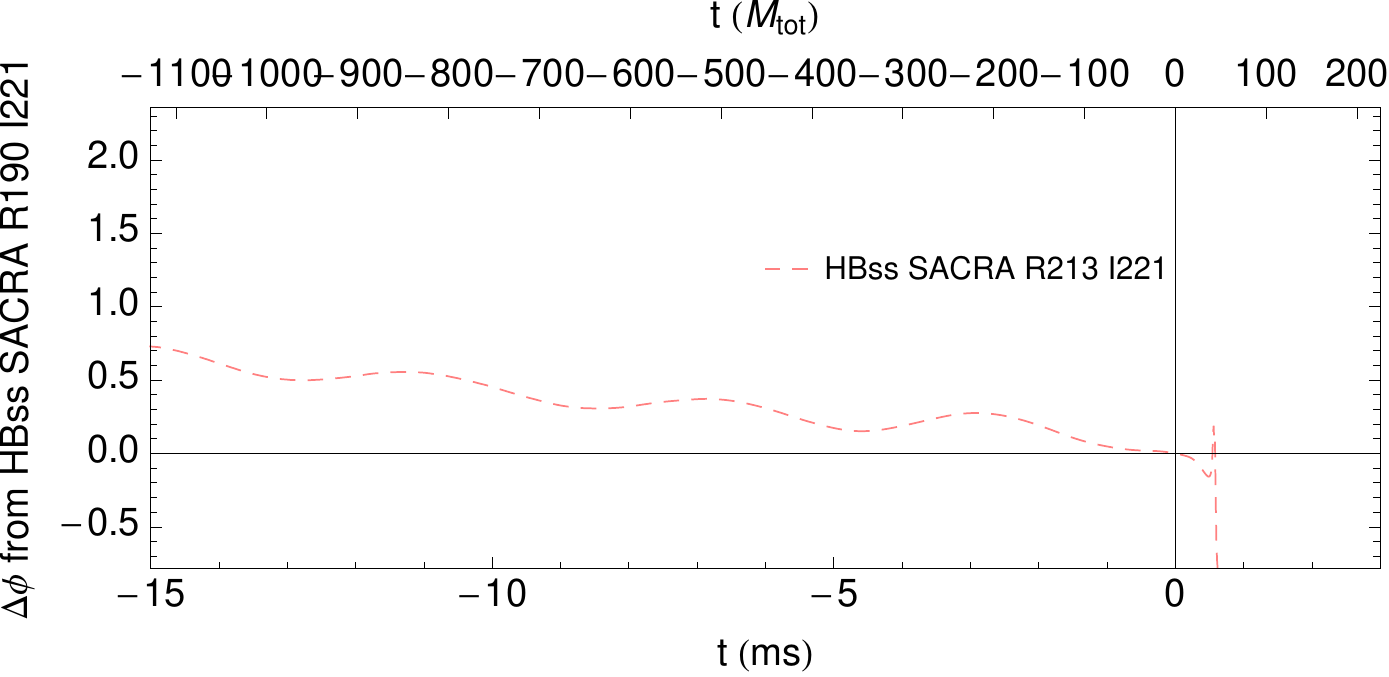} &
\includegraphics[width=8.5cm]{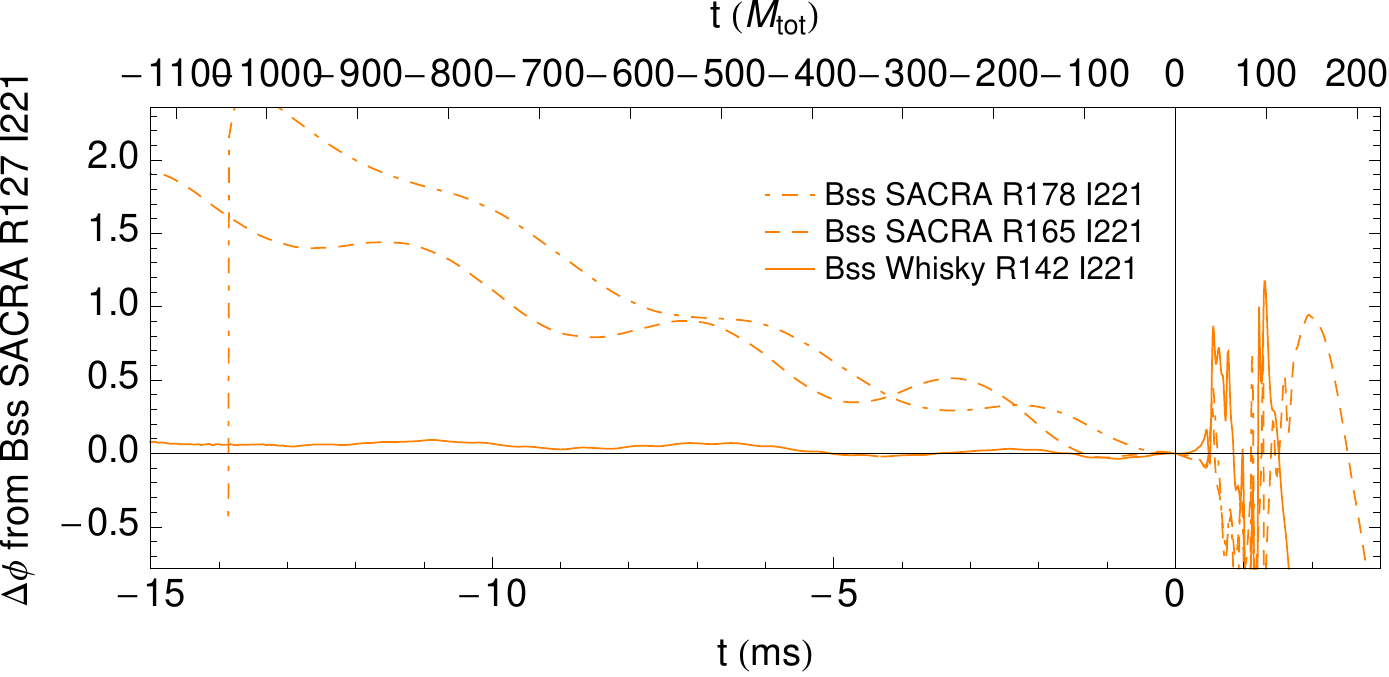} \\
\end{tabular}
\caption{Accumulation of phase differences between numerical simulations at
different resolutions and initial frequencies, relative to a reference
waveform for each EOS\@. We fix $t=0$ and $\phi=\arg h=0$ for all
waveforms at the peak amplitude point (see text). Note that the convergence of
the waveforms is poor after the peak amplitude (when the two stars begin to merge).
\label{fig:deltaphimerg}}
\end{figure*}

We will align multiple waveforms with the same EOS so that all waveforms have
the same time and phase when they reach their maximum amplitude
(Fig.~\ref{fig:mergalign} and Fig.~\ref{fig:deltaphimerg}),  which also allows
the comparison of waveforms with different initial separations.  The numerical
waveforms have residual oscillations in their amplitude as they approach the
peak, so we smooth this by taking a moving average of the amplitude over a range
of 0.5\,ms before finding the maximum amplitude. 

Different simulations of the same physical system, including those with
differing initial data, agree well through the last orbits when the waveforms
are compared this way, which we consider a strong indication that the dynamical
phase is being reliably simulated. We then estimate numerical inaccuracies on
the waveforms relative to the peak amplitude time to determine how much of the
inspiral we will use in subsequent analysis.

Looking at a set of waveforms, we find a common structure that is seen for each
EOS in Fig.~\ref{fig:mergalign}. As the neutron stars spiral toward each other,
at some point there is a transition from an inspiral phase to a merger or
coalescence phase, indicated by a maximum in the amplitude at the end of the
inspiral phase.  The retarded time of the peak amplitude corresponds roughly to
the impact of the two stars, after which shocks form and thermal and other
effects are expected to contribute to the waveform~\cite{STU03,STU05,ST06,
OJ07,HKOSK}.

\begin{figure}[tb]
\includegraphics[width=8.5cm]{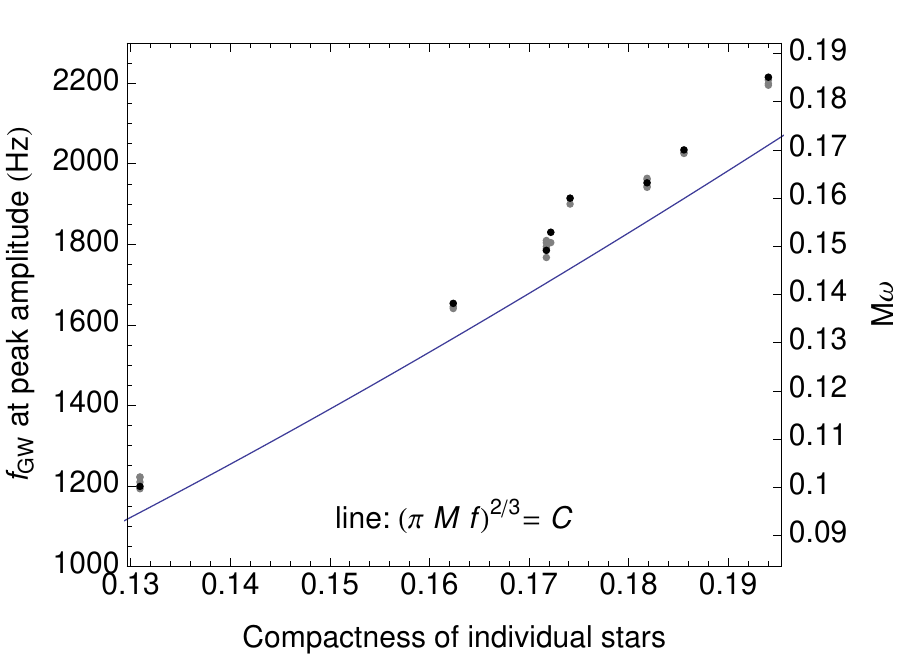}
\includegraphics[width=8.5cm]{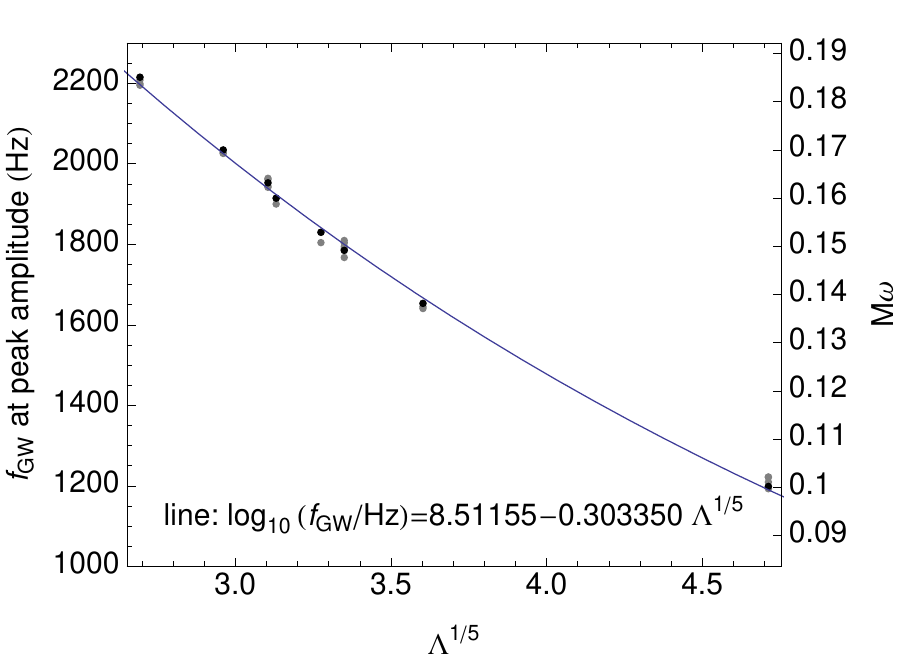}
\caption{Instantaneous gravitational-wave frequency at the point of peak
amplitude, as a function of the tidal parameter $\Lambda^{1/5}$ (bottom panel)
and as a function of individual star compactness $C$ (top). For each model, the
highest-resolution simulation for a given EOS is plotted in black,
lower-resolution simulations in grey.
 The $x=\left(\pi M f\right)^{2/3}=C$ relation used in
\cite{Damour2012} to characterize merger frequency is shown in the
compactness plot. An empirical fit using $\Lambda^{1/5}$ is shown in the bottom
plot; the frequency of merger is more tightly correlated with $\Lambda$ than
with compactness/radius.\label{fig:mergefreq} } \end{figure}

Somewhat surprisingly, we find that the parameter $\Lambda$ effectively
characterizes properties at this peak amplitude. Fig.~\ref{fig:mergefreq} shows
the frequency at peak amplitude as a function of both compactness, the
dimensionless ratio $M/R$ for an individual star, and $\Lambda$ for the
individual stars. We find that the frequency varies more smoothly with
$\Lambda$ than with radius or compactness; a linear fit of $\log f_{\text{GW}}$ as a
function of $\Lambda^{1/5}$ is displayed.  This may be an analogous relation to
those explored in \cite{Yagi2013}.

The instantaneous frequency of the gravitational waveform continues to
increase for a short time after the peak amplitude is reached, as the stars
coalesce.  A minimum in the gravitational wave amplitude follows, around which
the instantaneous frequency is not well defined and may spike upwards or
downwards. 

After this point, the qualitative waveform behavior depends
strongly on the EOS: For EOSs with higher pressure at the relevant densities,
a differentially rotating hypermassive
object may be supported, producing a quasi-periodic post-merger oscillation
waveform~\cite{Baiotti08,Sekiguchi2011}. This can last for tens of
milliseconds before the remnant collapses to a black hole~\cite{Rezzolla:2010}.  EOSs with lower
pressure, conversely, collapse quickly to black holes and have short
post-coalescence signals at roughly ringdown frequency, though of lower
amplitude than their binary black hole counterparts. The exact frequency
and amplitude of this signal varies with the EOS\@.

The differences
in phase evolution of different waveforms with the same EOS are shown relative
to a well-resolved reference waveform in Fig.~\ref{fig:deltaphimerg}. We note
that the difference in the finest resolution between simulations explains much
of the phase difference for all EOS; difference between initial separation (and
resulting differences in eccentricity at merger) have relatively small effects
at these resolutions.

Furthermore, we note that the magnitude of the phase differences stemming from
difference in resolution depends on the EOS: more compact stars require higher
resolution to give comparably small phase error.  This makes a quantitative
comparison of resolution effects on waveforms with different EOSs more
challenging.  For this analysis, we use a simplified procedure: examining the
merger-aligned waveforms, we estimate which of the given waveforms are
sufficiently resolved by current simulations by comparison with the highest
resolution available.  We will consider only waveforms which differ from the
highest resolution simulation by less than 0.5\,radians over the last 15\,ms
before the peak amplitude is reached. The systematic error resulting from this
level of phase error is calculated for the various measurability estimates in
subsequent sections.

\subsubsection{Detectors considered and Fourier-amplitude spectra}

\begin{figure*}[tb]
\begin{tabular}{cc}
\includegraphics[width=9cm]{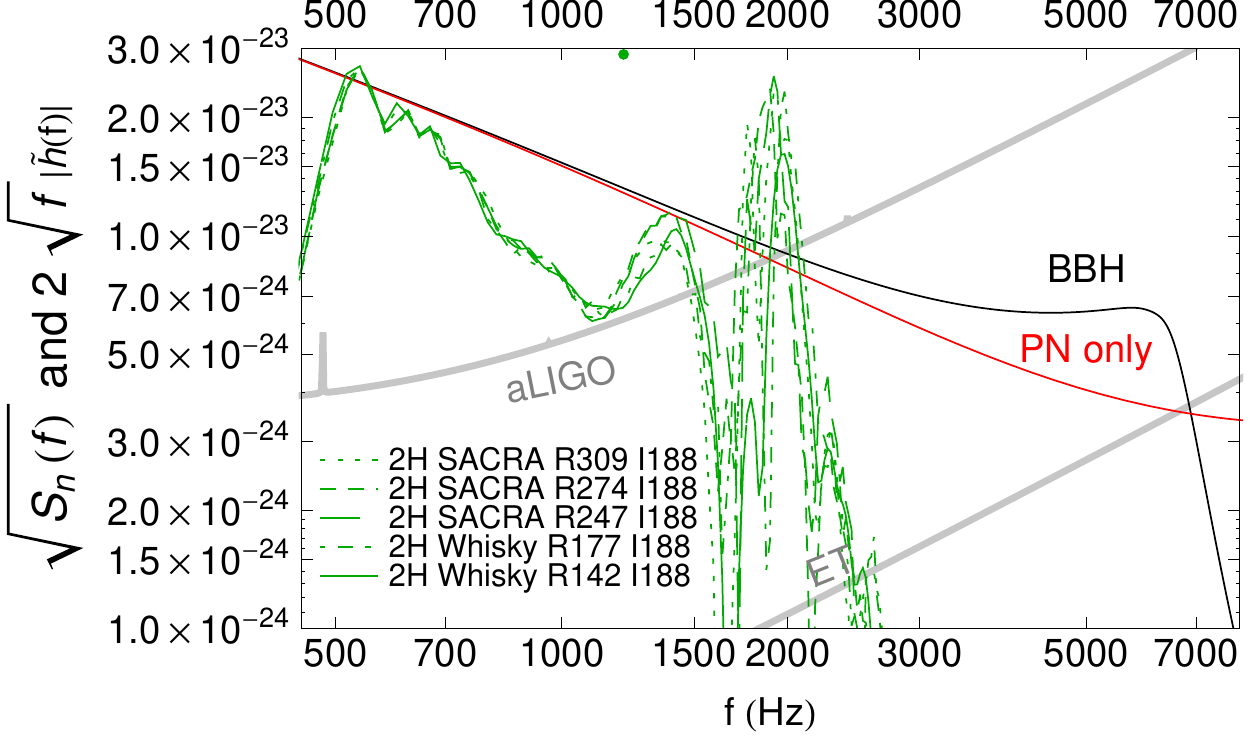} &
\includegraphics[width=9cm]{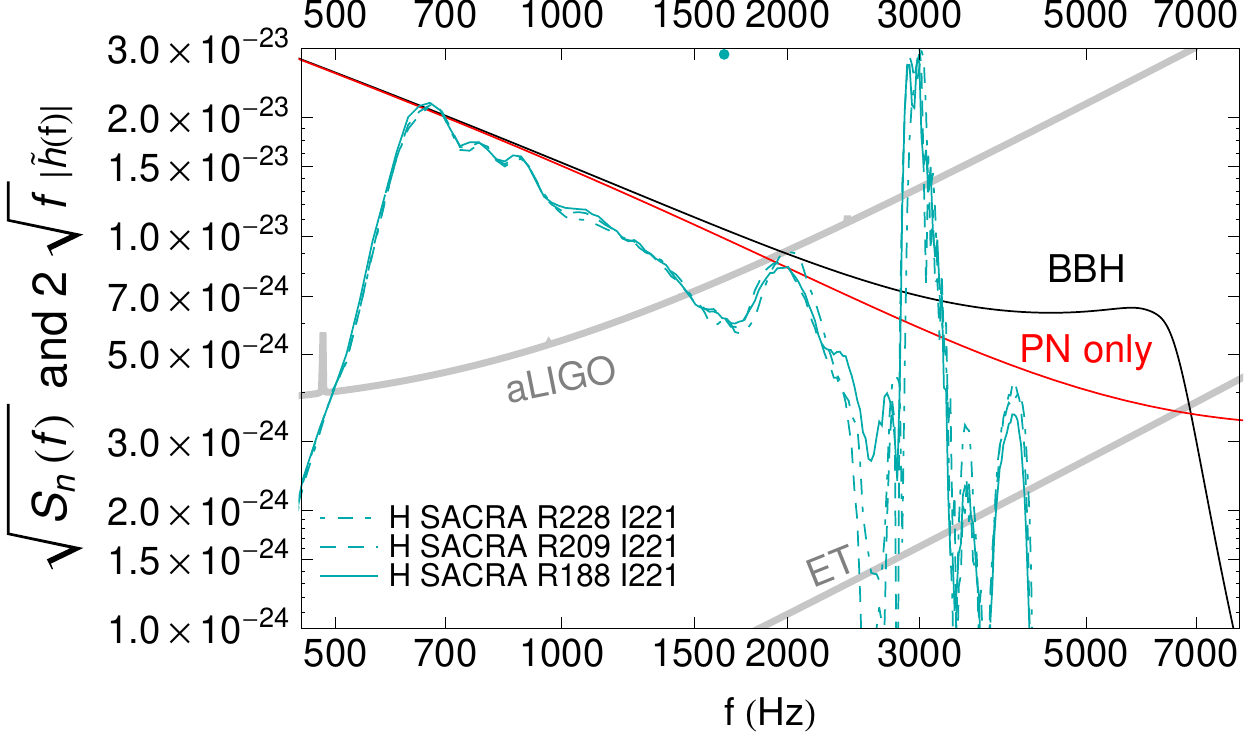} \\
\includegraphics[width=9cm]{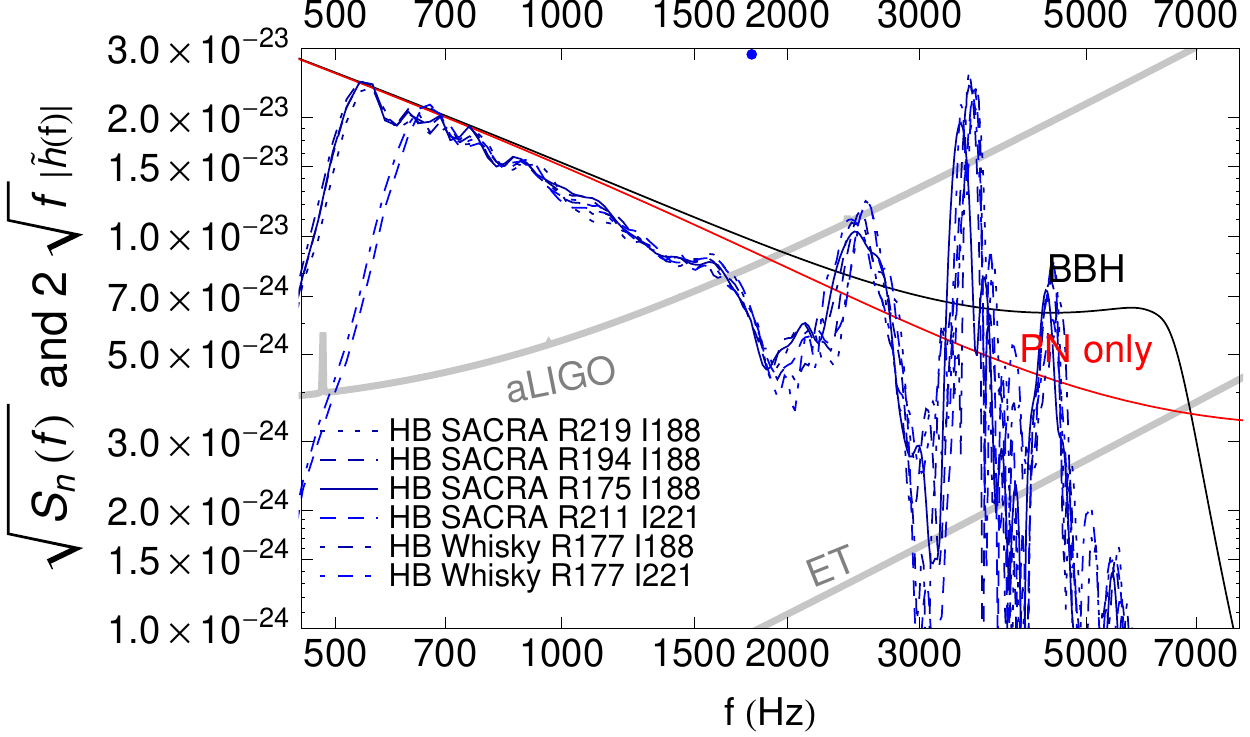} &
\includegraphics[width=9cm]{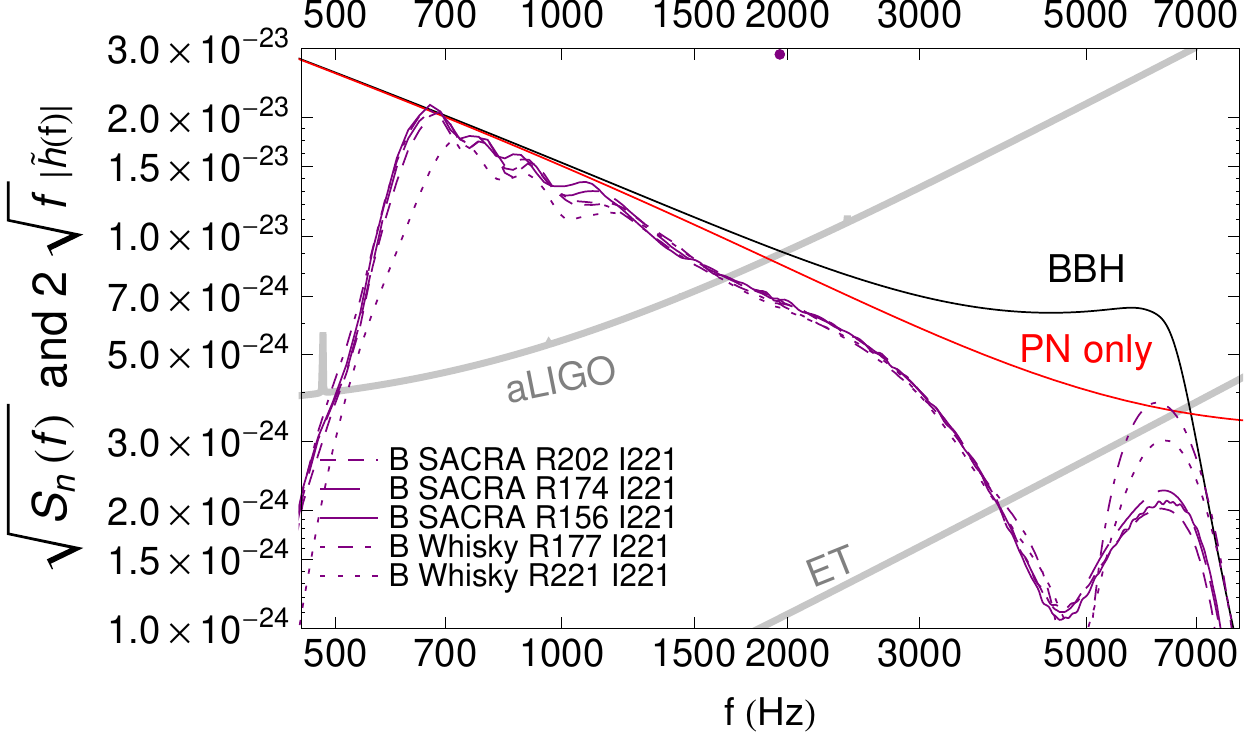} \\
\includegraphics[width=9cm]{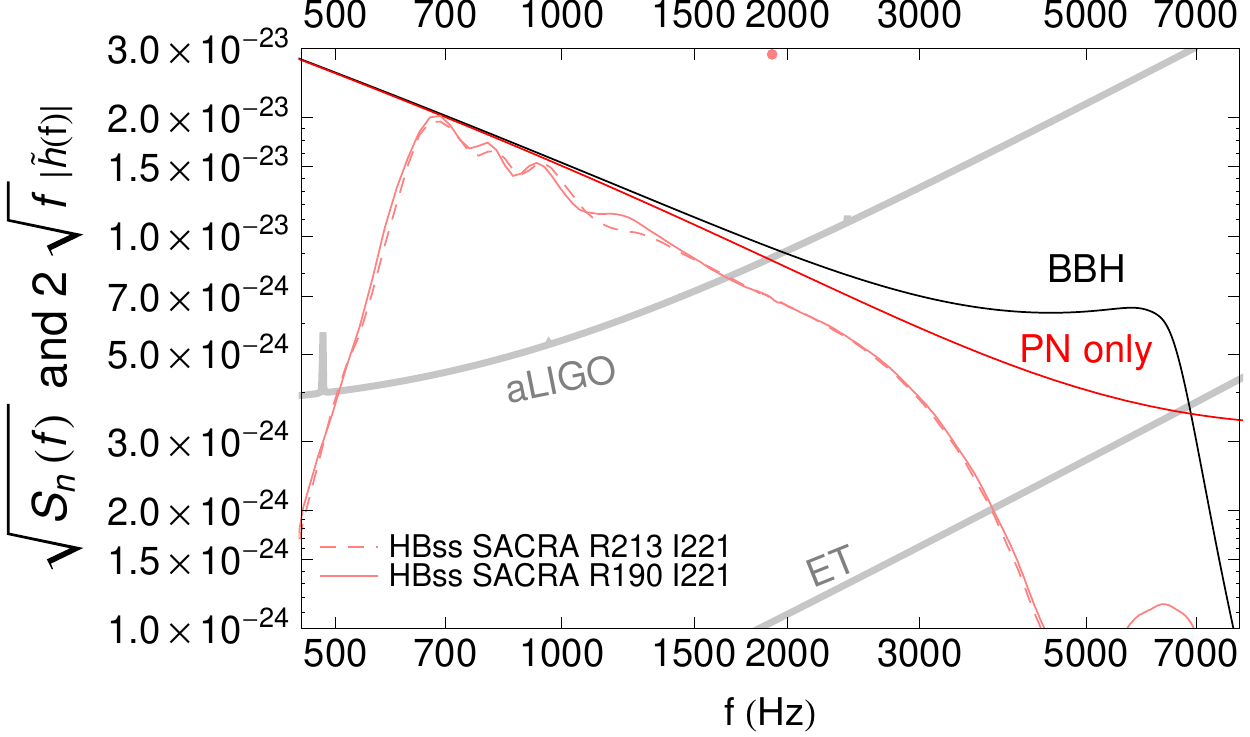} &
\includegraphics[width=9cm]{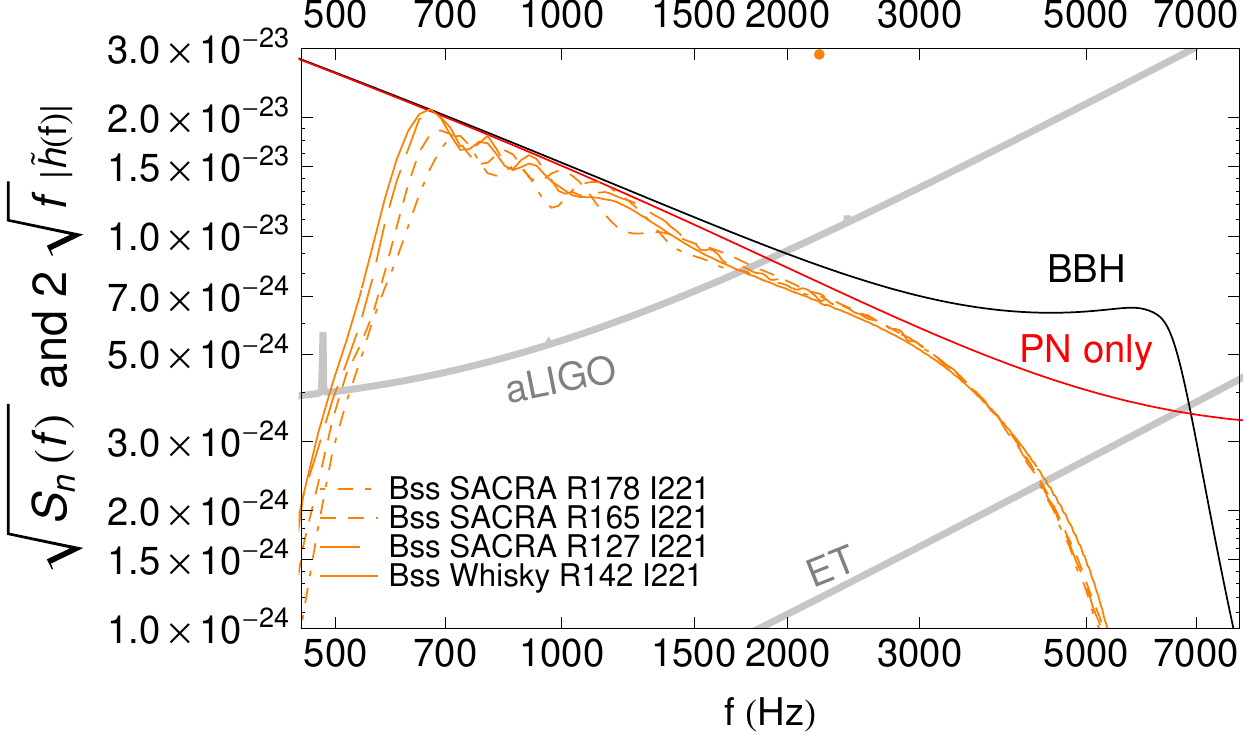} \\
\end{tabular}
\caption{Fourier spectra of numerical waveforms in units that facilitate the
comparison with gravitational-wave detector noise curves. Example noise spectra
are indicated by thick grey lines for the aLIGO high power noise~\cite{aLIGOnoise}
and the Einstein Telescope ET-D noise~\cite{Hild:2010id}. The starting
frequency depends on the initial orbital separation. The pre-merger waveform
gives a roughly monotonically decreasing amplitude, while post-merger
oscillations contribute spikes at high frequency ($1500$ Hz--$7000$ Hz). Black
curves indicate the phenomenological BH-BH waveform model
of~\citet{Santamaria2010} for the same mass parameters and red curves indicate
the stationary phase approximation of a pont-particle post-Newtonian inspiral.
The frequency of peak amplitude is
indicated by a colored dot on the upper axis.  }\label{fig:fourier}
\end{figure*}

For the merger of binary neutron stars, only detector configurations with good
high-frequency sensitivity will give useful constraints; broadband
configurations have previously been shown to compare favorably to narrowband
configurations tuned for high frequency sensitivity in distinguishing matter
effects~\cite{Read2009}. In this work, we choose the zero-detuning high-power
Advanced LIGO configuration~\cite{aLIGOnoise}, and the ET-D Einstein telescope
configuration~\cite{Hild:2010id}. 

We use a reference effective distance of $D_{\text{eff}}=100\,\text{Mpc}$ to
present results in this paper.  The effective distance $D_\text{eff}$ of a
binary system is same as the true distance of the system if it is optimally
oriented (face-on) and optimally located (directly above or below the detector)
and is greater than the true distance otherwise.  The amplitude of a signal is
inversely proportional to its effective distance.

The rate of signals with $D_\text{eff}=100$\,Mpc or smaller can be estimated by
comparing this to the fiducial Advanced LIGO horizon distance---the effective
distance of a detectable signal---of 445\,Mpc for NS-NS inspirals~\cite{Abadie:2010cf}. Within this
horizon, we expect roughly 40 (0.4--400) detectable inspirals per
year. Since rate scales with $D_\text{eff}^3$ for
sufficiently large distances, we expect $1\%$ of the detected signals to be as
strong or stronger than our reference signal, making it a plausible ``loudest''
signal over a few years of observation with realistic event rates. However, we
will also consider how our results scale to other values of $D_\text{eff}$ and the
constraints that weaker signals would place on the EOS in the sections to
follow.  

The estimates in this paper conservatively use only a single Advanced LIGO
detector. However, two detectors are being upgraded in the United States~\cite{AdvancedLIGO}, a
third is planned in India~\cite{LIGO-India}, and an upgrade of comparable
high-frequency sensitivity is underway for Virgo~\cite{AVIRGO} in Italy;
finally, a Japanese detector, KAGRA~\cite{KAGRA} is under construction,
although KAGRA's sensitivity curve is shifted slightly to lower frequency.
A multiple-detector network will provide additional discriminatory power,
reducing the statistical (though not the systematic) errors from those
estimated in this work.

In Fig.~\ref{fig:fourier}, we compare the amplitude of the simulated waveforms
as a function of frequency for each EOS to the strain sensitivity of the
detectors. Results of simulations with different codes, resolutions, and
initial separations are overlaid for each physically distinct inspiral. The
amplitude of the Fourier transform is an incomplete representation of the
waveform; similarities in amplitude do not necessarily reflect similarities in
phase evolution and can camouflage slow secular phase contributions that
decohere two waveforms. However, the amplitude has the advantage of being
independent of shifts in time and phase between two waveforms. 

The consistent change of the spectra as $\Lambda$ (and radius) increase
shows the effect of the EOS on the waveform at high frequency.  While insufficient
resolution (e.g., the dotted EOS B curve) may result in artificially low
amplitudes at lower frequencies, varying resolutions tend to agree in amplitude
before the systems transition to merger the characteristic frequency of the
peak amplitude.  Note that the spectra of more compact neutron stars (EOS B,
HBss, and Bss) follow black-hole inspiral to higher frequencies, but have
significantly different merger/ringdown amplitudes. 

The finite length of the numerical waveforms leads to a drop-off in the
amplitude at low frequencies. In the lower-resolution runs,
resolution-dependent dissipation in early cycles also results in a decrease in
the Fourier amplitude at lower frequencies relative to higher-resolution
waveforms with the same initial separation. 

\subsection{Post-merger oscillations}

Post-merger oscillations dominate the gravitational-wave emission from
hypermassive neutron-star remnant formed after the merger. They are stronger,
lower-frequency, and longer-lasting than the ringdown of a black hole formed in
prompt collapse.  The amplitude of the post-merger oscillation spectra are
shown in
Fig.~\ref{fig:minalign}. If strong enough, the high-frequency signals
could be independently detected by a search triggered by the inspiral, and
could constrain a combination of cold~\cite{Bauswein12} and
hot~\cite{Bauswein2010a} EOS\@. However, the SNR available in these post-merger oscillations is significantly
smaller than that of the numerically-simulated inspirals in the detectors
considered,
as summarized in Table~\ref{tab:pmo}; we present $\rho
\times(D_{\text{eff}}/100\,\text{Mpc})$ with entries which equal $\rho$ at
$D_\text{eff}=100$\,Mpc, and note that $\rho$ scales as $1/D_\text{eff}$.

Results from this paper and others suggest that post-merger oscillations will
be more challenging to measure than the EOS effects on late inspiral and
merger. Although a hot oscillating remnant may persist for tens or hundreds of
cycles, our simulations show nonlinear coupling giving an effective damping
time of less than ten cycles until a low final amplitude is reached.  The use
of more realistic density-pressure relations, thermal effects, and 
magnetic-field amplifications may change significantly the longevity (and
thereby the spectral amplitude) of these signals.  If we model the post-merger as a damped
oscillation of a single frequency, the SNR will scale roughly as $\tau^{1/2}$
for longer-lasting oscillations \cite{Creighton1999}. The post-merger oscillations
in the current simulations have multiple overtones as seen in
\cite{Stergioulas2011}, which spread the SNR over a range of frequencies, and
produce the oscillations in instantaneous frequency after merger in
Fig.~\ref{fig:mergalign} for EOS H and HB\@. They display a roughly exponential
decay in amplitude $A\sim \text{exp}\left(-t/\tau\right)$ over timescales
$\tau=3$\,ms to $6$\,ms. 

\begin{table}[tb]
\caption{SNR of post-merger waveforms in advanced
detectors, and approximate peak frequency $f_p$ of the oscillations.  Cases 2H,
H, and HB, show post-merger oscillations from a hypermassive remnant, and the 
roughly exponential decay timescale of the post-merger oscillations is shown. In
other cases, the neutron stars collapse to a black hole promptly after merger,
with suppressed ringdown. The spectra can
be seen in Fig~\ref{fig:minalign} \label{tab:pmo}} 
\begin{ruledtabular}
\catcode`?=\active
\def?{\kern\signwidth}
\begin{tabular}{lcccc}
 & \multicolumn{2}{c}{$\rho \times(D_{\text{eff}}/100\,\text{Mpc})$} &
 $f_{p}$& $t_\text{decay} $\\ \cline{2-3} EOS & aLIGO Broadband & ET-D  &
(kHz)& (ms)\\ \hline 
2H & 0.75--0.91 & 6.4--7.8 & 1.8 & 3--6\\
H & 0.54--0.57 & 4.5--4.7 & 3.0 &4--5 \\
HB & 0.43--0.47 & 3.5--3.9& 3.5 &3--4 \\
B & 0.04--0.07 & 0.4--0.6 & 6.5--7& \\
Bs & 0.04--0.06 & 0.3--0.6 &6.5--7& \\
Bss &0.03--0.06 &0.3--0.6 &6.5--7&\\
HBs &0.04--0.06 & 0.4--0.5 & 6.5--7&\\
HBss & 0.04--0.05& 0.4--0.5 &6.5--7&\\
\end{tabular}
\end{ruledtabular}
\end{table}

\section{Measurability using only numerical results}
\label{sec:measure}

Ideally, a data analysis program would coherently combine information from the
numerical waveforms (valid at high frequencies) with post-Newtonian waveforms
incorporating tidal effects (valid at low frequencies).  Joining a numerical
waveform to a theoretical post-Newtonian waveform relies on extremely accurate
numerical simulations with very large initial orbital simulation, as well as an
inspiral model that captures all relevant effects up to and including the
matching region.  While we will attempt this hybridization procedure in
Sec.~\ref{sec:hybmeas}, we begin with a simpler approach. 

If we assume that the low-frequency theoretical waveform correctly measures the
mass parameters and effective distance of the components, but cannot be
coherently combined with a numerically simulated waveform at higher
frequencies, we can still use the numerical simulations to try to identify the
EOS that best reproduces the high-frequency evolution, without using
the information about $t_0$ and $\phi_0$ measured from the low-frequency
waveform. The numerical waveforms must all be allowed to shift in time and
phase individually to find the best match to the observed gravitational-wave
data: the parameters $t_0$ and $\phi_0$ are marginalized over when measuring
the tidal effects.  

\subsection{EOS-based differences in numerical waveforms}\label{sec:numonly}

We first consider whether differences between EOS are significant in this
scenario.  We restrict ourselves to considering only the inspiral part of
the waveform, before the stars merge, where the cold EOS is expected to be an
accurate description of relevant physics and the numerical results are
convergent. To cut off the post-merger portion of the waveforms smoothly, the
natural minimum in amplitude (as shown in Fig.~\ref{fig:mergalign}) is taken as
the truncation point after each inspiral. 

\begin{figure}[tb]
\includegraphics[width=8.5cm]{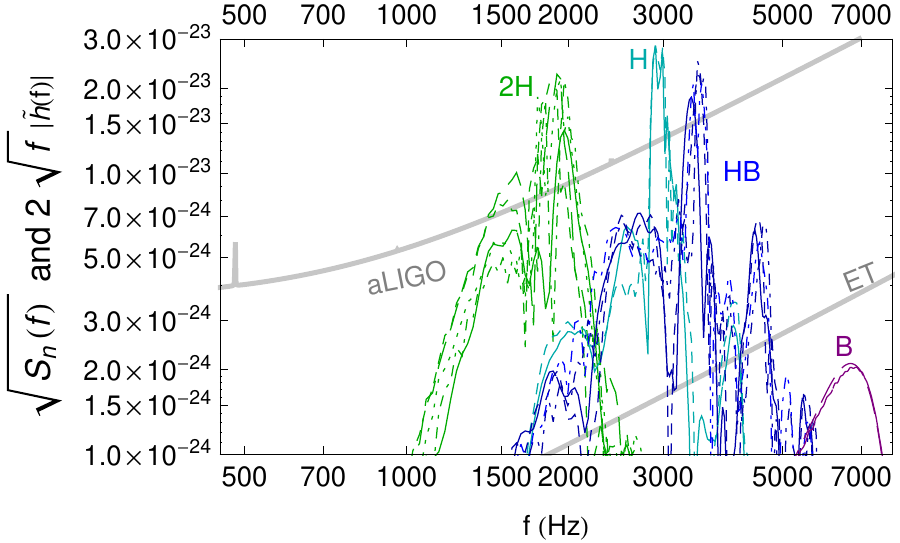}
\includegraphics[width=8.5cm]{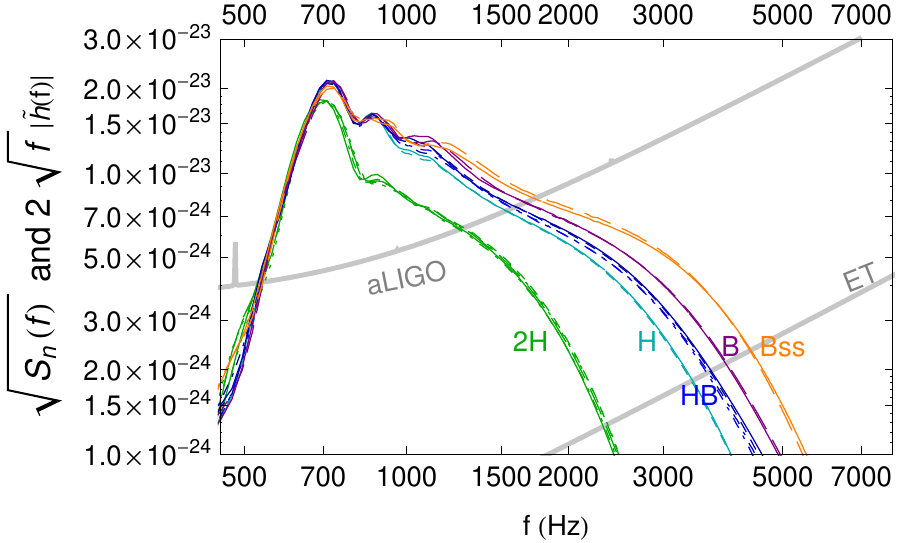} \caption{Top panel:
Post-merger waveforms for the different EOS, with lines as described in
Fig~\ref{fig:mergalign}. Bottom panel: Numerical inspiral-to-merger templates
as described in Sec.~\ref{sec:numonly},
which are smoothly turned on at 600Hz and stop at the minimum following the
peak amplitude.\label{fig:minalign} } \end{figure}

Since our waveforms began with varying initial separation, and some residual
effect of initial data can be expected at early times, we drop the portion of
the time-domain waveforms before a fixed instantaneous frequency.  To do this
consistently, the instantaneous frequency is first averaged over segments of
1.5\,ms to reduce residual eccentricity effects, and then a one-sided Hann
window  of width 4\,ms, centered on the time where the averaged frequency
reaches 600\,Hz, is applied to the waveform data. Similar windowing was used in
\cite{Read2009}.  Fourier-domain amplitudes of the resulting numerical inspiral
templates are shown in the bottom panel of Fig~\ref{fig:minalign}.

\subsection{Distinguishability}\label{sec:meas}

We wish to estimate our ability to distinguish between waveforms from different
numerical simulations, given a detected signal of the appropriate mass parameters.

To determine what model waveform best characterizes a detected signal,
we make use of the noise-weighted inner
product.  This inner product of two waveforms $h_1$ and $h_2$, for a detector 
with noise spectrum $S_h(f)$, is defined by
\begin{equation}
\langle h_1 \mid h_2 \rangle \equiv 4 {\mathrm{Re}} \int_{0}^{\infty}
\frac{\tilde{h}_1(f)\tilde{h}_2^*(f)}{S_h(f)}df.
\end{equation}
In terms of this inner product, the characteristic signal-to-noise ratio of a
given waveform $h$ is $\rho \equiv \langle h \mid h \rangle^{1/2} $.

Two waveforms, $h_1$ and $h_2$ are said to be marginally
\emph{distinguishable} if the quantity
\begin{equation}
\label{eq:deltarho}
\|\delta h\| \equiv \|h_2 - h_1\| \equiv \sqrt{\langle h_2 - h_1\mid h_2 - h_1 \rangle}
\end{equation}
has a value $\|\delta h\|
\gtrsim1$~\cite{steve,Lindblom:2008cm,Read2009,MacDonald2011}.

We wish to consider the minimum value of $\|\delta h\|$  over all possible
relative shifts in time and phase between the template waveforms, and it turns out
to be most efficient to calculate this via the overlap between two waveforms.
With the complex waveform $h$ constructed for this analysis,
and methods similar to \citet{FINDCHIRP} and \citet{EffectiveFisher}, we use
the inverse Fourier transform appropriate to $\tilde{h}$ to construct a complex
overlap as a function of timeshift $\tau$ for each polarization: 
\begin{equation}
\langle h_{1\times,+} (t +\tau) \mid h_2 (t) \rangle \equiv 4 
\int_{0}^{\infty} \frac{\tilde{h}_{1\times,+}(f)\tilde{h}_2^*(f)}{S_h(f)}\text{e}^{2\pi i
f \tau} df.
\end{equation} 
The absolute value of this quantity at a given $\tau$ is the maximum overlap
possible with shifts in phase. Maximizing its absolute value over $\tau$ thus
gives the maximum overlap for arbitrary shifts in both time and phase. 

\begin{table}[tb]
\caption{The first row shows the expected SNR
$\times(D_{\text{eff}}/100\,\text{Mpc})$ 
of the numerical inspiral-to-merger
waveforms described in Sec.~\ref{sec:numonly}, for each EOS\@. Note that the
signal's presence, amplitude, and mass parameters are assumed to be established
from an inspiral detection.  Subsequent rows show the expected SNR of
\emph{differences} between these waveforms and waveforms of the row-labelling
EOS, minimized over shifts in time and phase. The SNRs are calculated for each
possible pair of resolved waveforms, and the mean and standard deviation of the
resulting estimates for each pair of EOS are tabled. 
\label{tab:rhodiff}} 
\centering
Advanced LIGO
high-power detuned
\smallskip
\begin{ruledtabular}
\catcode`?=\active
\def?{\kern\digitwidth}
\begin{tabular}{lccccc}
EOS&2H & H& HB	 & B & Bss\\
\hline
SNR &2.22 &2.77 &2.81 &2.87 &2.89\\
\hline
2H& 0.10$\pm$0.08 & 1.85$\pm$0.02 & 1.93$\pm$0.04 & 2.02$\pm$0.03 & 2.03$\pm$0.02 \\
H&  & 0.06$\pm$0.06 & 0.66$\pm$0.06 & 1.03$\pm$0.06 & 1.13$\pm$0.03 \\
HB& & & 0.09$\pm$0.06 & 0.61$\pm$0.07 & 0.86$\pm$0.03 \\
B&  &(symm.) &  & 0.11$\pm$0.13 & 0.52$\pm$0.06 \\
Bss&  &  &  &  & 0.06$\pm$0.07 \\
\end{tabular}
\end{ruledtabular}
\bigskip
Einstein Telescope configuration D
\smallskip
\begin{ruledtabular}
\catcode`?=\active
\def?{\kern\digitwidth}
\begin{tabular}{lccccc}
EOS&2H&	 H	 & HB	 & B&Bss	\\
\hline
SNR &22.3 &27.4 &27.8 &28.2 &28.4\\
\hline
2H& 1.1$\pm$0.9 & 17.4$\pm$0.3 & 18.2$\pm$0.4 & 18.9$\pm$0.3 & 18.9$\pm$0.2 \\
H& & 0.6$\pm$0.6 & 6.0$\pm$0.5 & 9.1$\pm$0.5 & 10.0$\pm$0.2 \\
HB& & & 0.9$\pm$0.6 & 5.5$\pm$0.7 & 7.5$\pm$0.3 \\
B& & (symm.) & & 1.2$\pm$1.3 & 4.6$\pm$0.6 \\
Bss& & & & & 0.6$\pm$0.7 \\
\end{tabular}
\end{ruledtabular}
\end{table}
\begin{figure}[tb]
\includegraphics[width=8.5cm]{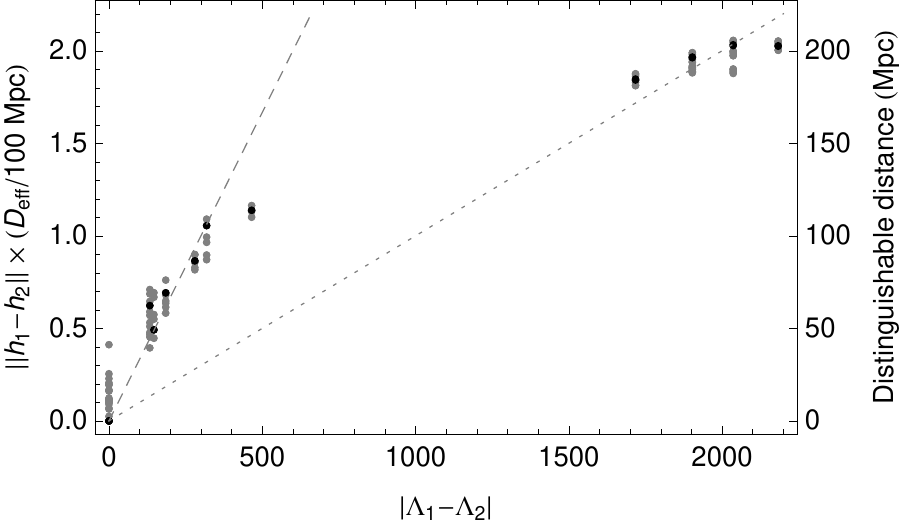} \caption{ For
numerical merger templates, $\|\delta h \|= \|h_1 - h_2\|$ between two waveforms is plotted as a function of $|\Lambda_1 - \Lambda_2|$ after being 
minimized over relative shifts in time and phase. The distance $D_\text{eff}$
at which two waveforms would be distinguishable is labelled on the right axis.
 The
result is not linear in $\Delta\Lambda$. 
At the reference
$D_\text{eff}=100$\, \mpc, the difference
between waveforms has $\|\delta h\|=1$ for $\Delta\Lambda\simeq500$ (dashed line)
and $\|\delta h\|=2$ for $\Delta\Lambda\simeq2000$ (dotted line). This plot
superimposes $\Delta\Lambda$ for all pairs of simulations for Advaned LIGO 
high-power zero-detuning and optimally-oriented systems at 100 Mpc; ET-D gives
similar plot with $\|\delta h\|$ increased by a factor of 10.  \label{fig:diffsnrplot}}
\end{figure} 

We use this maximum overlap to estimate the signal to noise ratio of the
difference between two templates
\begin{equation}\|\delta h\| ^2 \simeq \langle h_1 | h_1 \rangle + \langle
h_2 | h_2 \rangle - 2 \langle h_1 | h_2
\rangle_\text{max} 
\end{equation} 
where $\langle h_1 |
h_2\rangle_\text{max}$ is maximized over
shifts in time and phase.  Note that we do not normalize our templates: the
inspiral
detection is expected to
determine the relative amplitude expected at merger, and EOS which merge
earlier give real differences in expected SNR which will affect the maximum
likelihood, as can be seen in Table~\ref{tab:rhodiff}. Because the
value of $\|\delta h\|$ depends on the distance to the
signal, we record $\|\delta h\| \times(D_{\text{eff}}/100\,\text{Mpc})$.

The differences between waveforms are presented in Table~\ref{tab:rhodiff} for
the numerical waveforms discussed in Sec.~\ref{sec:numonly}.  EOS 2H, with the
largest difference from other EOSs (relative to EOS H,
$\Delta R = 2.95$\,km and $\Delta\Lambda=1717$), and produces a 
$\|\delta h \| \simeq2$.  The more realistic EOS give smaller differences, but H
and B, with $\Delta R = 1.3$\,km and $\Delta \Lambda=319$, are marginally
distinguishable at $D_\text{eff}=100$\,Mpc.  For a given pair of waveforms, we
can determine the maximum effective distance to which they can be
distinguished, where $\|\delta h \| =1$, since
$\|\delta h\|$ scales as
$1/D_\text{eff}$. The result is plotted as function of $\Delta \Lambda$ in
Fig.~\ref{fig:diffsnrplot}. 

Using numerical simulations that extend to earlier frequencies can increase the
distinguishability of EOS: numerical waveforms starting at orbital angular
frequency of $188$ are available for EOS 2H and HB, and can be used to
construct templates starting at $500$\,Hz, which have SNRs in Advanced LIGO of
$\rho_\text{2H}=3.24$ and $\rho_\text{HB}=3.61$ at the reference
$D_\text{eff}=100$\,Mpc. The resulting $\|\delta h \|=2.14\pm0.05$ is larger
than the $\|\delta h \|=1.93\pm0.04$ of templates starting at $600$\,Hz.
However, measures of systematic error roughly double; templates starting at
$500$\,Hz have more than twice the duration of templates starting at $600$\,Hz.
The relative impact of differing EOSs also becomes smaller at
earlier times---$\|\delta h\|/\rho$ is decreasing---and required
computational time will increase rapidly. Simulations will also be more challenging
for compact neutron stars, which require higher resolution for
equivalent accuracy.

The importance of numerical effects can be estimated in two ways: the
value of $\|\delta h\|_\text{syst}$ between two different waveforms for
the same EOS, and the variance in $\|\delta h\|$ between two EOSs that arises
from making different choices of the representative numerical waveform for each
EOS\@.  These results are included in \ref{tab:rhodiff} and are visible in the
spread of points at $|\Lambda_1 - \Lambda_2|=0$ in Fig.~\ref{fig:diffsnrplot}.
The $\|\delta h\|$ between two EOS changes by less than $\sim10$\% with
different waveform choices; however, while $\|\delta h\|_\text{syst}$ (between
numerical waveforms of the same EOS) is typically $0.1$ (or 10\%) at
$D_\text{eff}=100$\,Mpc, it reaches $0.4$ in the worst case.

\subsection{Parameter estimation}\label{sec:par}

Given a parameterized family of waveforms, $h(p_i)$, where $p_i$ includes an
EOS-dependent parameter of interest, we determine the value of the parameters
$p_i$ that produce the best match by comparing the detected signal to the
members of this family.
If the detected signal is $s$, then the most likely values for the parameters $p_i$
are those that best fit the data by minimizing the distance to the signal with
the above-defined inner product, $\left\langle s - h(p_i) | s - h(p_i)
\right\rangle$.  
For normalized templates, 
the best-fit $p_i$ maximize the \emph{overlap} $\left\langle s | h(p)
\right\rangle$ between signal and waveform family. 

The best-fitting values of $p_i$ will differ from the true values because
of two effects: The first effect is that the measured $p_i$ will be
shifted away from their true value because of the presence of random detector
noise; we describe this \emph{statistical} error by the root-mean-squared value
of the
parameter shift, $\delta p_{i,\text{stat}}$.  The second effect arises if there
is a fundamental difference between the true gravitational waveform and the
nearest member of the family of waveforms that are being used; such a
\emph{systematic} error is given by $\delta
p_{i,\text{syst}}~$\cite{CutlerVallisneri}.  The statistical error depends on the
amplitude of the signal relative to the level of detector noise, so it scales
inversely with the signal's SNR\@. The systematic error is SNR-independent.

For large SNR signals, 
the statistical error $\delta p_{\text{stat}}$ can be calculated using
the Fisher matrix formalism.  If a waveform is parameterized by a set of
parameters $\{p_i\}$, then the Fisher matrix is given by
\begin{equation} \Gamma^{ij} = \biggl\langle \frac{\partial
h}{\partial p_i} \,\bigg|\, \frac{\partial h}{\partial p_j}
\biggr\rangle, \end{equation}
and the statistical error associated with the measurement of a single parameter
$\lambda_j$ is
\begin{equation} \delta p_{j,\,\text{stat}} = \sqrt{(\Gamma^{-1})_{jj}},
\end{equation}
where the matrix $(\Gamma^{-1})_{ij}$ is the inverse of the Fisher matrix
$\Gamma^{ij}$~\cite{Finn:1992wt}. 

For the subsequent analysis, we will consider variation only in the single
parameter $\Lambda$ which best characterizes the EOS\@.
When restricting to cases where multiple well-resolved waveforms are available,
current simulations do not cover a two-dimensional region of the EOS parameter
space, so we are restricted to single-parameter estimates. For an example of
generalization to multi-parameter descriptions of the EOS, see~\cite{LKSBF}.

 The numerical simulations considered here are of equal-mass systems with fixed
total mass, so correlations with mass parameters can not yet be determined. We
assume that accurate measurements of the chirp mass $\mathcal{M}$ and
dimensionless mass ratio $\eta$ can be made from the detected inspiral
preceding the merger, which is a reasonable assumption for loud signals ($\rho
\gtrsim 20$).  In the post-Newtonian case, one can
(at least to first order) recast the tidal effect
of generic-mass systems in terms of a single effective
$\tilde{\Lambda}(\mathcal{M},\eta)$ of the system
\cite{Flanagan2008,Hinderer2010} which depends primarily on $\mathcal{M}$, and 
uncertainties in mass ratio do not overwhelm tidal effects. This may not be so
straightforward for the coalescence, especially if amplified
tidal disruption occurs in unequal mass systems. The dependance of such a
$\tilde{\Lambda}$ on the less-easily measured mass ratio parameter may also
obfuscate the tidal dependence in more general systems. While spin should also
be considered in a full analysis, especially as it may obscure mass ratio
measurements, spin uncertainty has a relatively weak impact on measurement of
tidal parameters in binary neutron star systems \cite{Damour2012}.  Results in
the mixed binary case \cite{Lackey2013} suggest a factor of 3 increase in
$\delta\Lambda$ when phenomenological inspiral-to-merger waveforms are used for
a coherent analysis of both mass and tidal parameters, compared to an analysis
considering tidal variation alone. 

Given a single discretely sampled parameter,
the Fisher ``matrix'' can be estimated using a 
finite difference approximation to the derivative as
$\Gamma\approx\|\delta h \|/(\Delta \Lambda)^2$.
This finite difference estimate of the random error in a parameter is then
given by~\cite{Read2009} 
\begin{equation}
\label{eq:deltaprand}
\delta \Lambda_{\text{rand}} = \frac{| \Lambda_1 - \Lambda_2| }{
\sqrt{\langle h_1 - h_2 | h_1 - h_2 \rangle}},
\end{equation}
or $\delta \Lambda_{\text{rand}} = \Delta\Lambda / \|\delta h\|$, and we
note that the $\delta \Lambda_\text{rand}$ is exactly the value of $\Delta
\Lambda$ where two waveforms are distinguishable using the criteria above.

While Eq.~(\ref{eq:deltaprand}) is an approximation to the Fisher matrix, it is
a more accurate characterization of the information contained in a finite
strength signal than the Fisher matrix itself, as discussed in \citet{EffectiveFisher} and references
therein. The usefulness of the Fisher matrix breaks down in part when 
overlap between two signals of $\Lambda_1$ and $\Lambda_2$ no longer scales
quadratically with $\Lambda_1-\Lambda_2$. We can directly calculate 
the overlap between the two signals to determine how well two parameter
values can be distinguished. In the Fisher matrix analysis, a first-order and
linear relation $\|\delta h \| \propto \Delta \Lambda$ is assumed, as
would be valid for small $\Delta \Lambda$; in the discrete approximation,
nonlinear structure in $\|\delta h \| $ as a function of $\Delta \Lambda$ is
revealed (Fig.~\ref{fig:diffsnrplot}).  The estimate of $\delta
\Lambda_\text{rand}$ will thus depend on the effective SNR scale. The
\emph{effective} $\delta \Lambda_{\text{rand}}$ gives $\|\delta h\|=1$ for
for marginally distinguishable signals.  The difference in $\Lambda$ that is
marginally distinguishable at a given $D_\text{eff}$ is
the expected $\delta\Lambda_\text{rand}$ at that distance.

When $\Lambda$ is used to parameterize the simulations, we find that plotting
$\|\delta h\| = \|h_2-h_1\|$ versus $\Delta\Lambda=|\Lambda_2-\Lambda_1|$ for all
available choices of $\Lambda_1$ and $\Lambda_2$ gives a well-defined
pattern: $\|\delta h \|$ is well-described as a function of $\Delta\Lambda$
only, so there is only weak dependence of
$\delta\Lambda_\text{rand}$ on the value of $\Lambda$. This is not true for
other parameter choices, such as $\Lambda^{1/5}$ or radius. For the
$D_\text{eff}=100$\,Mpc reference waveform here, $\delta
\Lambda_\text{rand} \simeq 300$, but for a $D_\text{eff}=200$\,Mpc signal
$\delta\Lambda_\text{rand} \simeq 2000$.

If the true signal waveform $g$ differs from all members of the parameterized
family of waveforms $h(\{p_i\})$ then there will be a systematic error
in the measurement of the parameters $\{p_i\}$\cite{CutlerVallisneri}; the systematic error
is given by
\begin{equation}
  \delta p_{j,\,\text{syst}}
  = \sum_i (\Gamma^{-1})_{ij} \biggl\langle h-g \,\bigg|\,
  \frac{\partial h}{\partial p_i} \biggr\rangle.
\end{equation}
To assess the systematic error associated with imperfections in the numerical
waveforms, we take $g$ and $h$ to be variant waveforms that  purport to represent
the same system, e.g., numerical waveforms from two simulations of the same
EOS, and as before we replace the derivative with respect to the parameter with
a finite difference of waveforms with different EOS parameters.  With the subscript labeling the choice of EOS, the resulting
approximate formula for the systematic error in measuring the parameter 
$\Lambda$ is
\begin{equation}\label{eq:delta-syst}
\delta \Lambda_{\text{syst}} \approx ( \Lambda_1 - \Lambda_2 ) \frac{\langle
h_1 - g_1 \mid h_1 - h_2 \rangle}{\langle
h_1 - h_2 \mid h_1 - h_2 \rangle}.
\end{equation}
If we apply a Cauchy-Schwarz inequality to the numerator of the above equation, we find that 
\begin{equation}
|\delta \Lambda_{\text{syst}}| \lesssim \left| \Lambda_1 - \Lambda_2 \right|
\left( \frac{\langle
h_1 - g_1 \mid h_1 - g_1 \rangle}{\langle
h_1 - h_2 \mid h_1 - h_2 \rangle}\right)^{1/2},
\end{equation}
which we can rewrite using magnitude of the difference between the two variant
waveforms, $\|\delta h \|_\text{syst}$, compared to the magnitude of the difference
of two waveforms of different parameter values $\|\delta h \|$ defined in
Eq.~(\ref{eq:deltarho}), as 
\begin{equation}
|\delta \Lambda_{\text{syst}}| \lesssim |\Delta \Lambda|
\frac{\|\delta h\|_\text{syst}}{\|\delta h\| }.
\end{equation}
Both of the $\| \delta h \|$ scale with effective distance, giving a constant
$\|\delta h\|_\text{syst}$ for a given $\Delta\Lambda$.  In an effective error
calculation, where the estimate appropriate to a given $D_\text{eff}$
is the value of $\Delta\Lambda$ where $\|\delta h\|=1$, the relative
systematic error will simply be the ratio of $\|\delta h\|_\text{syst}$ to
$\|\delta h\|$ at that distance. 

Later we will present estimates of systematic error 
$\|\delta h\|_\text{syst} \times(D_\text{eff}/100\,\mpc)$. In Table
\ref{tab:rhodiff}, the diagonal entries are the average $\|\delta h\|_\text{syst}$ at the reference distance. In plots such as Fig.~\ref{fig:diffsnrplot} and \ref{fig:diffsnrplothyb}, these
are visible as the scatter of points above $|\Lambda_1-\Lambda_2|=0$.

\section{Hybrid construction and improved measurability}
\label{sec:hybmeas}

For low-mass binary systems, such as those which include neutron stars,
numerical waveforms start at frequencies that are high compared to the
sensitive band.  Ideally, EOS effects will be measured using hybrid
waveforms which combine post-Newtonian inspiral (including tidal effects) with
the numerical simulation results.  However this introduces additional sources
of systematic error, as discussed in~\cite{Hannam2010a,MacDonald2011}.  If a numerical simulation is begun at too high a
frequency, the theoretical point-particle post-Newtonian (or other analytical
inspiral waveform) will no longer be valid.  If the resolution of the numerical
simulation is too low (so that there is too much numerical dissipation through
the high-frequency band in which the tidal effects become strong) then reliable
hybrid waveforms cannot be constructed. 

In this work, we use results of highly accurate numerical simulations
\cite{Bernuzzi2011,Hotokezaka2013} to justify extending our analytic model to
sufficiently high frequencies that the simulations considered in this work,
which cover more EOS parameter space with lower resolution, will have
sufficient accuracy to model the final orbits without systematic error
overwhelming our estimate.  

\subsection{Hybrid construction}\label{ss:hybconst}

We fix a baseline 3.5 order post-Newtonian Taylor-T4 model~\cite{BBK07,DIS01}
for subsequent analysis.  While the impact of choosing a post-Newtonian
expansion is large in the last orbits, this choice accurately mimics equal-mass
binary black holes up to $M\omega=0.01$~\cite{BBK07}, which is within $114M$ of
peak amplitude for all binary neutron stars simulated here: the hybrid
waveforms do not use the post-Newtonian inspiral waveform beyond the frequency range where it
approximates binary black holes.

We include post-Newtonian estimates of the tidal contributions to the waveform
phasing from~\cite{Flanagan2008,Vines:2011ud}, at leading and next-to-leading
order, which have been shown to give potentially significant contributions to
measurability for the EOS considered if waveform models are extended to high
frequency~\cite{Hinderer2010,Damour2012}.

Our inclusion of tidal effects is done by simply adding
additional contributions to the baseline model. In a full parameter estimation,
a calibrated phenomenological or EOB description of the point-particle dynamics
may be required to accurately capture intermediate-order post-Newtonian terms.
However, \cite{Damour2012} shows that the magnitude of tidal phase contributions
in EOB is accurately approximated by the addition of post-Newtonian tidal terms
into 2.5 or higher post-Newtonian models, justifying their use of Taylor F2
waveforms for measurability estimates. Here we add Newtonian and post-Newtonian
tidal contributions to the 3.5 post-Newtonian Taylor T4 waveforms used to model
point-particle dynamics; the differences between waveforms of different EOS
should likewise be accurately captured by this scheme.

It has been conjectured~\cite{Damour2010,Baiotti2010} that additional
higher-order post-Newtonian tidal corrections would be required to match of the
inspiral of numerical waveforms, but the calculated next-to-leading order terms
in \cite{Vines:2011ud} were smaller than those previously obtained by
fitting~\cite{Baiotti2011}. More recently, \cite{Bernuzzi2012} and
\cite{Hotokezaka2013} have each independently calculated the expected waveforms
of binary neutron star inspiral and merger using high-resolution numerical
simulations with careful error analysis. The phase evolution of these
high-accuracy waveforms is compared to both post-Newtonian and EOB waveforms,
which each incorporate current analytically calculated tidal terms. Within
numerical uncertainties, both groups find that the numerical waveforms and the
various analytic inspiral models all agree until roughly $300M$ to $500M$
before merger. 

In this work, we do not have waveforms with the same level of accuracy, but
we restrict our analysis to use only the last 15\,ms or $1128M$, of the
numerical waveforms, over which the effect of the waveform resolution used in
this work is small for the ``resolved'' waveforms we have been considering
(Fig.~\ref{fig:deltaphimerg}). We use the high-accuracy waveform results to
justify our use of Taylor T4 inspiral models, with leading-order and
next-to-leading order tidal terms, for earlier times.

We note that, in addition to secular tidal effects, there may be other effects
that are not encompassed in the post-Newtonian (or current EOB) expansion
framework, for example f-mode resonances~\cite{Kokkotas1995,Stergioulas2011}.
Our transition to a numerical waveform in  the final orbits will include the
high-frequency contributions of such effects, but they are not incorporated
in the analytic model.

To construct hybrids, we match the analytic and numerical waveforms over a
time-domain matching region using the maximum correlation method
of~\cite{Read2009}. If one defines the
complex correlation $z$ in a restricted time domain
$\{T_{\text{I}},T_{\text{F}}\}$ for two waveforms $h_1(t)$ and
$h_2(t)$ with a relative time shift $\tau$ by 
\begin{equation}
z(\delta t;h_1,h_2)\equiv \int_{T_{\text{I}}}^{T_{\text{F}}}
h_{1}^{\vphantom{\ast}}(t) h_{2}^{\ast}(t-\tau) dt,
\end{equation}
then the correlation between the two waveforms with no phase shift is
$\Re z(\tau ;h_1,h_2)$. Introducing
a phase shift $\delta\phi$ to $h_{2}$ produces a correlation  $ \Re
\exp({\mathrm i}\delta\phi) z$. For a given $\tau$, the maximum correlation
between two waveforms for any phase shift will be  $|z|$, and the phase
shift which produces that correlation is $\delta\phi = -\arg z$ (cf.~\cite{FINDCHIRP}).

Previous waveform analyses have performed similar matching
via least squares difference over a segment~\cite{Hannam2009, Santamaria2010}
or have matched time and phase at a single point in the
inspiral~\cite{Boyle2009}, either in the time or frequency domain.
Our procedure maximizes a cross-term averaged between polarizations which
contributes negatively to the least squares distance between waveforms.  It is
a time-domain analogue of procedure used above to maximize Fourier-domain
overlap in Sec.~\ref{sec:meas}---for infinitely long time domains, it is
equivalent to an unweighted frequency-domain match, similar to that used for
detection.

\begin{figure}[tb]
\includegraphics[width=8.5cm, trim=1.3cm 0 1.6cm 0
,clip=true]{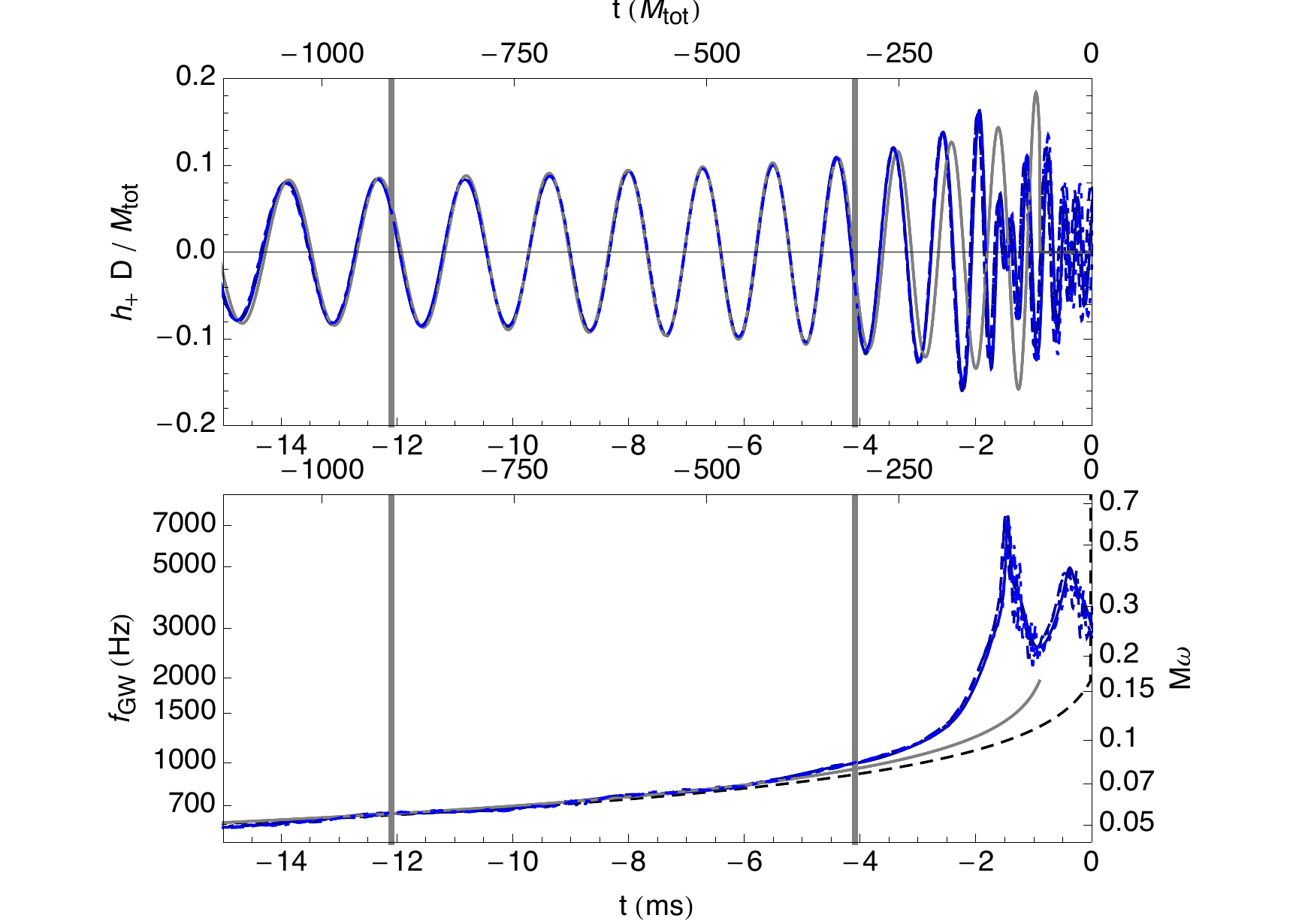} 
\caption{Example hybrid
construction for the four resolved waveforms with EOS HB\@. The reference $t=0$ is
the coalescence time of a point particle inspiral. A post-Newtonian inspiral
with tidal corrections appropriate to EOS HB is shown in grey. The numerical
waveforms used to construct hybrids overlaid with their maximum-correlation
alignment between vertical lines indicate the start and end of the numerical
matching region. The frequency dependence of the waves is also shown; the
post-Newtonian point particle inspiral is shown with a dashed black line, the
post-Newtonian with tidal corrections in grey, and the four numerical waveforms
in blue following the line indication scheme of Fig.~\ref{fig:mergalign}.
\label{fig:hybex}}
\end{figure}
\begin{figure}[tb]
\includegraphics[width=7cm]{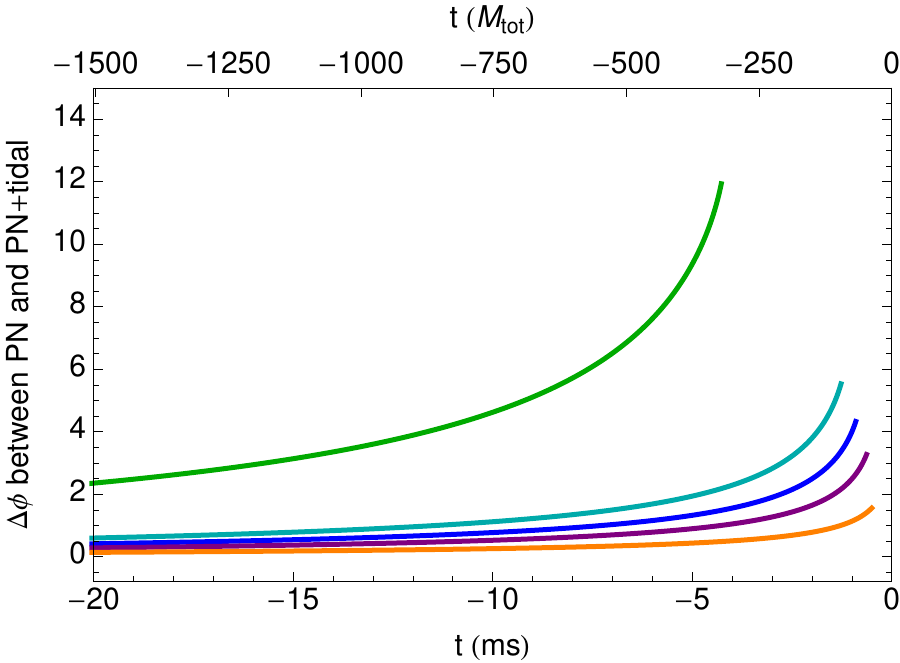}
\includegraphics[width=7cm]{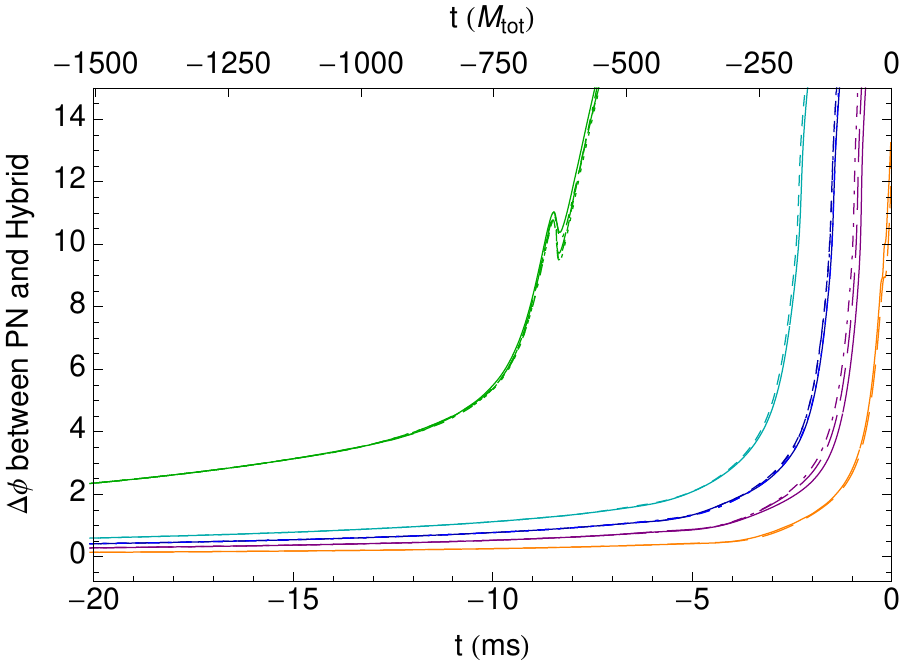}
\caption{Top panel: The phase departures from point-particle Taylor-T4 due to
post-Newtonian tidal contributions. From highest to lowest, the lines
indicate EOS 2H, H, HB, B, and Bss. Bottom panel: The phase departures due to
hybrid waveforms, with lines as described in Fig.~\ref{fig:mergalign}.
Integration is begun at 200\,Hz, after the accumulation of the majority of the
SNR \cite{Damour2012}
is expected to have fixed the relative phase, but before
significant tidal contributions arise.\label{fig:MergePhase}}
\end{figure}

We use only the final orbits and transition to merger from the numerical
waveforms, as captured in the final 10\,ms or $752M$ before merger:
specifically, the match region is set relative to
the time of peak amplitude for each numerical waveform, from $(t_\text{peak}-10$\,ms$)$ to
$(t_\text{peak}-2$\,ms$)$. The numerical waveform is aligned to post-Newtonian
inspiral by the maximum correlation above, and then a hybrid waveform is
constructed by windowing between inspiral and numerical waveforms over the last
half of the match region. An example of this construction is shown in
Fig.~\ref{fig:hybex} for the full set of variant HB waveforms. The resulting
hybrids are Fourier-transformed, and the amplitude of the difference between
two waveforms $\langle h_1-h_2 | h_1-h_2\rangle$ is calculated directly using
the inner product defined in Sec.~\ref{sec:meas}.

\begin{figure}[tb]
\includegraphics[width=8.5cm]{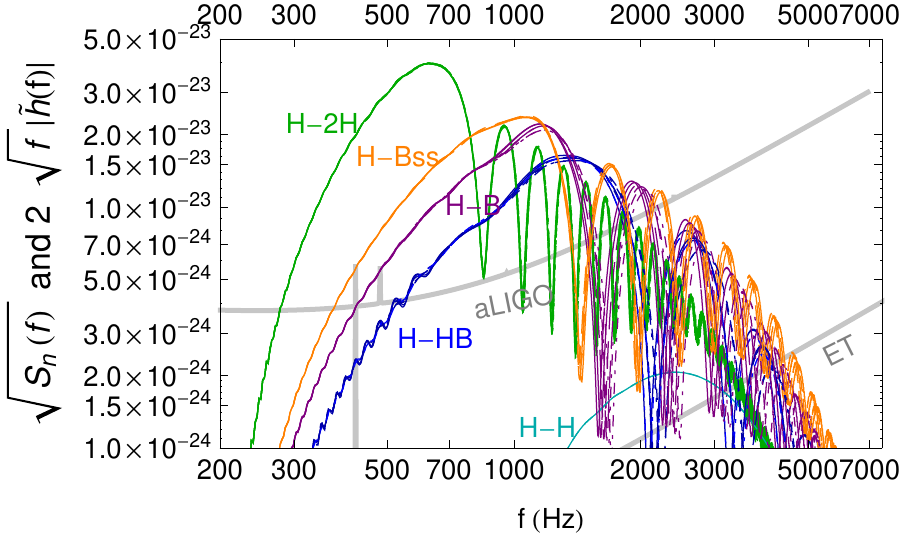}
\caption{The difference between H hybrid waveforms and waveforms with other
EOS is plotted relative to ET and Advanced LIGO noise curves, showing the
frequency range which produces the measurable difference.  EOS 2H has the
largest difference, seen in the spectrum labelled ``$\mbox{H}-\mbox{2H}$", followed by
Bss (``$\mbox{H}-\mbox{Bss}$"), B (``$\mbox{H}-\mbox{B}$"), and
HB (``$\mbox{H}-\mbox{HB}$")\@. Lines calculated with hybrids
constructed from different simulations lie roughly on top of each other, and
the differences between the two EOS H simulations (``$\mbox{H}-\mbox{H}$") lie substantially lower.  The
amplitude of the difference becomes larger than the amplitude of the component
hybrids when they are perfectly out of phase, and the oscillations at high
frequency show the two waveforms moving in and out of phase.
\label{fig:mergediff}} \end{figure}

\subsection{Measurement using hybrid waveforms}\label{ss:hybmeas}
To calculate the differences between hybrid EOS, we use a somewhat less
conservative estimate than in the previous sections; to save computational
time, we do not include the long low-frequency portion of the waveform in our
analysis and therefore cannot minimize differences over shifts in time and
phase.  However, the time and phase of a high-frequency waveform that is
coherent with the low-frequency inspiral are no longer free parameters. 
\citet{Damour2012} calculate the frequency range over which each waveform
parameter is determined: 90\% of the total $\rho^2$ from a binary neutron star
inspiral is collected from frequencies below 200\,Hz. Mass parameters
are determined using the waveform at even lower frequencies, and tidal effects
on the inspiral are determined only by the highest-frequency portion; the two
regions decouple.

 We assume that the waveform portion below $200$\,Hz, which is virtually
identical for models of different EOS, will fix the relative time and phase of
template and signal waveforms; if the overlap of very long PN-only waveforms
with different tidal contributions is maximized over variations in time and
phase, the relative time of coalescence is approximately that of waveforms
which are exactly aligned at $200$\,Hz.  We then
consider only differences that accumulate from $200$\,Hz and up when comparing
waveforms of different EOS. The resulting phase accumulation  is shown in
Fig.~\ref{fig:MergePhase}. With the contributions from the inspiral, the
differences $|h(\Lambda_1)-h(\Lambda_2)|$ between the hybrid waveforms become
more significant.  The SNR of the differences between all pairs of waveforms is
shown in Table~\ref{tab:rhodiffhyb}.  Fig.~\ref{fig:mergediff} illustrates
the Fourier transform of the difference between pairs of waveforms, in which
one member has EOS H, plotted as a signal against the Advanced LIGO noise
curve.

\begin{figure}[tb]
\includegraphics[width=8.5cm]{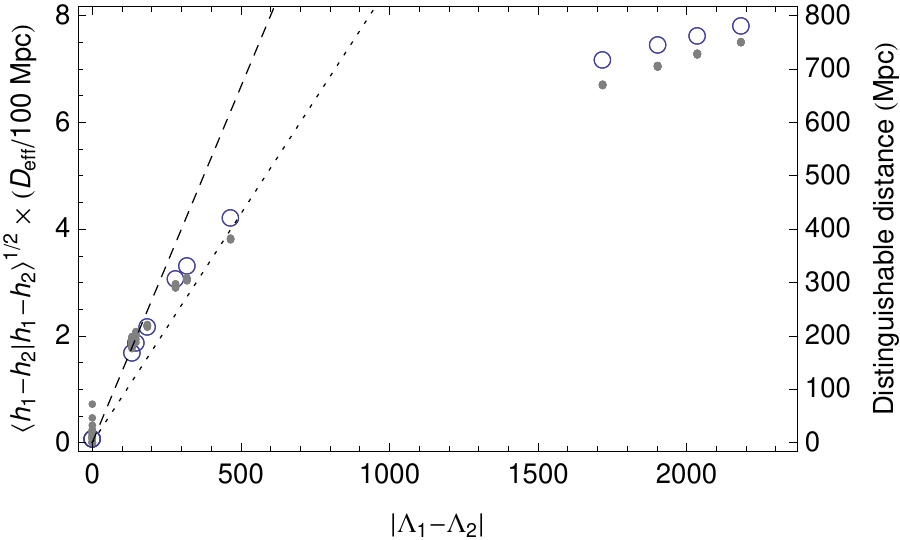} \caption{ For
hybrid waveforms aligned at $200$\,Hz, the distinguishability is estimated
using the inner product of differences between waveforms.  At the reference
$D_\text{eff}=100$\,\mpc, the difference between waveforms has
$\|\delta h\|\simeq2$ for $\Delta\Lambda=150$ (dashed line) and
$\|\delta h\|\simeq3$ for $\Delta\Lambda=350$ (dotted line).  These results are for
Advanced LIGO high-power zero-detuning; ET-D gives similar plot with an order
of magnitude increase in $\rho$ (and decrease in distinguishable distance).
Note that the distinguishability is improved by a factor of 3 to 4 compared to
numerical-only estimates in Fig.~\ref{fig:diffsnrplot}.
\label{fig:diffsnrplothyb}} \end{figure}

As before, we compile the set of differences for all waveform pairs into a plot
of $\|h_1 - h_2\|$ vs.~$|\Lambda_1-\Lambda_2|$ in
Fig.~\ref{fig:diffsnrplothyb}.  The result is again not linear in
$\Delta\Lambda$, so the statistical
error estimate will depend nonlinearly on the loudness of the signal. At the
reference $D_\text{eff}=100$\,Mpc, the difference between waveforms has
$\|\delta h\|\simeq2$ for $\Delta\Lambda=150$, allowing each EOS to be
distinguished from a binary black hole. Marginally distinguishable parameter
differences are at then $\delta\Lambda=150$ at $D_\text{eff}=200$ and
$\delta\Lambda=350$ at $300$\,Mpc (where BNS inspirals are
detected with
$\rho\simeq16$ and $11$). This means that a combination of weaker signals can be
used to give significant constraints on the EOS\@, as seen in \cite{delPozzo2013}.

\begin{table}[tb]
\caption{ The first row shows
SNR $\times(D_{\text{eff}}/100\,\text{Mpc})$ 
for the full hybrid waveforms. The remaining rows show 
 $\|\delta h\| \times(D_{\text{eff}}/100\,\text{Mpc})$ 
between hybrid waveforms, averaged over
resolved waveforms for each EOS\@. The standard deviation of the set
of resulting estimates is also provided. The average difference between
waveforms of the same EOS is a measure of systematic error from numerical
inaccuracies for the given hybridization procedure. \label{tab:rhodiffhyb}} \centering
Advanced LIGO
high-power detuned
\smallskip
\begin{ruledtabular}
\catcode`?=\active
\def?{\kern\digitwidth}
\begin{tabular}{lccccc}
EOS&2H & H& HB	 & B & Bss\\
\hline
SNR &33.72 & 33.78 & 33.78 & 33.79 & 33.80\\
\hline
2H&0.08$\pm$0.06 & 6.70$\pm$0.01 & 7.05$\pm$0.01 & 7.28$\pm$0.01 & 7.5$\pm$0.01 \\
H& & 0.08$\pm$0.10 & 2.18$\pm$0.02 & 3.06$\pm$0.03 & 3.82$\pm$0.01 \\
HB& & & 0.13$\pm$0.10 & 1.87$\pm$0.09 & 2.94$\pm$0.02 \\
B& &(symm.) & & 0.35$\pm$0.31 & 2.01$\pm$0.09 \\
Bss& & & && 0.07$\pm$0.08 \\
\end{tabular}
\end{ruledtabular}
\end{table}

\subsection{Additional systematics with hybridization}\label{ss:hybsyst}

The systematic error stemming from alternate methods of generating
the waveforms and alignment used to calculate measurability can be
estimated using Eq.~(\ref{eq:delta-syst}) where the two waveforms $g$ and $h$
may be taken to be different hybrid waveforms. As hybrid waveforms
incorporate choices in the construction beyond simply choosing the
numerical waveform, additional systematic errors are introduced. 

Systematic errors are estimated using the set of ``well-resolved'' waveforms,
using the criteria of Sec.~\ref{ss:peak}, which have phase differences of
$\sim0.1$ radians to $\sim0.4$ radians over the last $1.5$\,ms (1100$M$) before merger.
For a fixed hybrid construction method and PN model, the diagonal entries of
Table~\ref{tab:rhodiffhyb} give $\|\delta h|_\text{syst}\lesssim0.3$ at
$D_\text{eff}=100$\,Mpc, for $\|\delta h\|_\text{syst}/\|\delta h\|$ of roughly
5--20\% from variant numerical simulations of these EOS.

\begin{table} 
\caption{Effect of  shifting match window in hybrid
construction to earlier times with waveform resolutions used in this analysis:
$\|\delta h \|_\text{syst}$ at 100\,Mpc between two hybrids constructed with the same
numerical waveform, or between a hybrid waveform and a PN inspiral waveform.
Systematic errors decrease as waveform resolution increases; more compact stars
require higher resolution. For EOS Bss, the hybridization error is as large
as that from neglecting hybridization entirely; this can also be seen in the
difference between orange curves in Fig.~\ref{fig:radlambda}.
\label{tab:syshyb}}
\begin{ruledtabular}
\catcode`?=\active
\def?{\kern\digitwidth}
\begin{tabular}{lcccc}
& \multicolumn{2}{c}{Hybrid Variation}& \multicolumn{2}{c}{PN Inspiral}\\
\cline{2-3}\cline{4-5}
EOS & aLIGO & ET-D & aLIGO & ET-D\\
\hline
 2H \sacra R309 I188   & 1.60 & 14.86 & 2.34 & 22.08 \\
 2H \sacra R274 I188   & 1.48 & 13.73 & 2.34 & 22.13 \\
 2H \sacra R247 I188   & 1.46 & 13.52 & 2.35 & 22.22 \\
 2H \whisky R177 I188  & 1.16 & 10.63 & 2.34 & 22.09 \\
 2H \whisky R142 I188  & 1.08 & 09.90 & 2.35 & 22.22 \\
 H \sacra R209 I221    & 0.97 & 08.57 & 1.75 & 15.71 \\
 H \sacra R188 I221    & 0.87 & 07.66 & 1.75 & 15.62 \\
 HB \sacra R194 I188   & 0.88 & 07.67 & 1.68 & 14.90 \\
 HB \sacra R175 I188   & 0.89 & 07.76 & 1.62 & 14.40 \\
 HB \whisky R177 I188  & 1.09 & 09.53 & 1.65 & 14.63 \\
 HB \whisky R177 I221  & 0.93 & 08.15 & 1.63 & 14.44 \\
 B \sacra R174 I221    & 0.90 & 07.79 & 1.58 & 13.87 \\
 B \sacra R156 I221    & 0.85 & 07.41 & 1.42 & 12.48 \\
 B \whisky R177 I221   & 1.32 & 11.53 & 1.62 & 14.26 \\
 Bss \sacra R127 I221  & 1.36 & 11.80 & 1.48 & 12.88 \\
 Bss \whisky R142 I221 & 1.47 & 12.83 & 1.48 & 12.90 \\
\end{tabular}
\end{ruledtabular}
\end{table}

We explore the impact of changing hybridization procedures by shifting the
window used to match PN and numerical waveforms, within the assumptions outlined in Sec.~\ref{ss:hybconst}. The procedure of
Sec.~\ref{ss:hybmeas} is repeated with a variant match window of
$(t_\text{peak}-12$\,ms$)$ to $(t_\text{peak}-4$\,ms$)$. The results for
$\|\delta h\|$, distinguishability, and measurability of $\Lambda$ do not
change appreciably. However, the systematic error from differences between
numerical simulations doubles as
earlier inspiral portions of lower-resolution numerical waveforms come into
play. 

We estimate the impact of uncertainty in the hybridization
procedure used to produce parameter-estimation templates by comparing waveforms
constructed from the same numerical simulation using different hybridization
windows. The results in Table \ref{tab:syshyb} show that the variant
hybridization gives $\|\delta h\|_\text{syst}\simeq0.9$--$1.6$, which decreases
with increasing resolution. Even with the best resolution,
$\|\delta h\|_\text{syst}$ ranges from 20--75\% of $\|\delta h\|$ at 100\,Mpc,
largest for compact stars and small
differences in EOS.  As in the binary black hole case \cite{MacDonald2011},
longer and more accurate numerical simulations will be required to reduce the
systematic error associated with hybridization.  We also note that while these
variants seem to cover a reasonable  range given the assumptions outlined in
Sec.~\ref{ss:hybconst}, a more systematic analysis with accurate waveforms
would be required to quantify the uncertainties for parameter estimation.

The tidal contributions discussed in this paper include the leading order and
next-to-leading order tidal contributions from~\cite{VF2010,VFH2011}. The
significance of higher-order PN tidal terms can be estimated by dropping the
next-to-leading order tidal contribution. This results in systematic error of
$\approx 9$--$15\%$ of $\|\delta h \|$ at 100\,Mpc; always smaller than
systematic error from varying the hybrid procedure, but most important for
large-radius stars.

We have throughout assumed that an underlying point-particle inspiral model is
accurate up to $M\omega\simeq0.1$. In practice, for equal-mass systems, the
time-domain Taylor-T4 signal appears to satisfy this requirement, but
calibrated phenomenological or EOB models may be required to accurately capture
the underlying dynamics of more general systems.  We also assume, based on the
agreement seen in \cite{Bernuzzi2012,Hotokezaka2013}, that there
are no EOS effects beyond tidal contributions before the hybridization times
used in this paper; this neglects any contributions smaller than best current
numerical errors and low-frequency resonances \cite{Balachandran2007}. 

\section{Use of inspiral-only templates} 
We can also estimate the impact of neglecting numerical simulation results
entirely in a waveform model of binary neutron star inspiral and merger. To do
this, we calculate the difference between the hybrid waveforms and our inspiral
model extended to coalescence. The total $\|\delta h\|_\text{syst}$ is shown in Table
\ref{tab:syshyb}. For compact stars (EOS B, Bss), the hybrid error is
comparable to the error from neglecting the numerical merger entirely. However,
larger neutron stars (EOS 2H, H) have a reduction in systematic error
from using hybrids instead of inspiral-only waveforms.
\begin{figure}[tb]
\includegraphics[width=8.5cm]{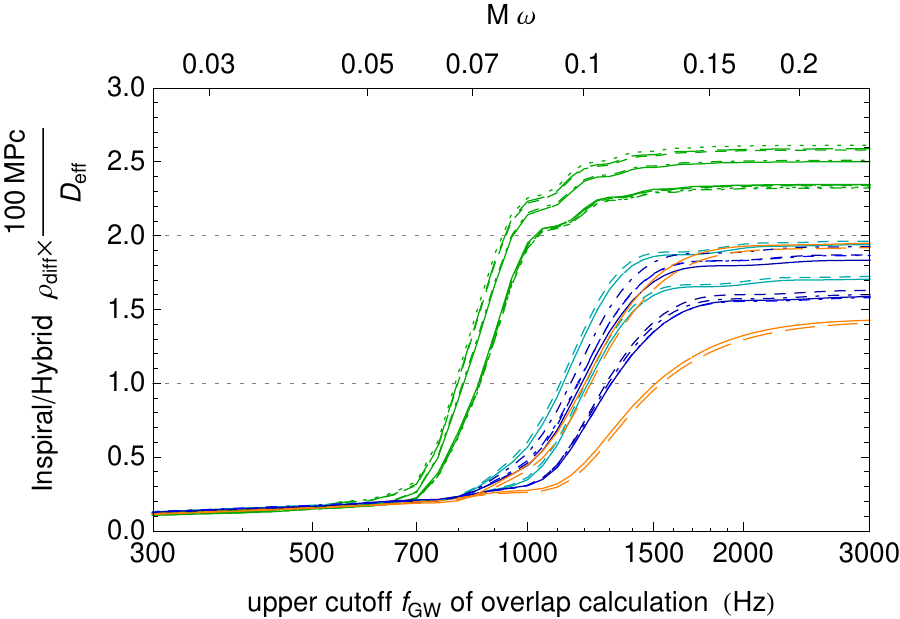}
\caption{The accumulation of $\|\delta h\|_\text{syst}$ between
inspiral-only waveforms and the two variant cases of  hybrid waveforms as a
function of upper cutoff frequency on the inner product. The impact of
hybridization is significantly larger for compact EOS (EOS Bss, in
orange, has a radius of 10.2\,km)---the total acumulation is comparable to the
difference between hybrids, as seen in Table \ref{tab:syshyb}.\label{fig:insphybdiff}}
\end{figure}

One can estimate the extent to which inspiral-only waveforms can be trusted by
considering an inner product calculated only up to an upper cutoff frequency.
The frequency dependence of the difference between hybrid and inspiral-only
waveforms is shown in Fig.~\ref{fig:insphybdiff}. If this cutoff frequency is
low, or the signal is weak, there is no measurable impact from using
inspiral-only waveforms. All hybrids and analytic
inspirals agree (to within numerical error) below 700\,Hz. However, strong
signals or large neutron stars produce significant differences from
post-Newtonian models. The EOS 2H model, which is an extremely large ($R=15.2$\,km)
neutron star, begins to depart from post-Newtonian inspiral at approximately
700\,Hz, even if the hybrid window includes higher frequencies,  and hybrids
constructed for EOS 2H reach $\|\delta h\|\times
\left(100\,\mpc/D_\text{eff}\right)=2.0$ (distinguishable at
$D_\text{eff}=2000\,\mpc$ with total $\rho\simeq16$) at $f\simeq1023$\,Hz.
Hybrids for more realistic EOS (H and HB) are not distinguishable from
post-Newtonian inspiral until total SNR $\rho\simeq22$ and upper frequency
$\sim1400$\,Hz to $\sim1600$\,Hz, although this is sensitive to the choice of
hybridization window.
 
Aside from the systematic error introduced by using inspiral-only waveforms to
measure EOS effects, one can consider the usefulness of inspiral-only models to
estimate EOS measurability.  In Fig.~\ref{fig:diffsnrplothyb}, we overlay the
result of an analogous estimation using only post-Newtonian inspirals,
including leading order and next-to-leading order tidal effects, and extended
to post-Newtonian coalescence. We again again align waveforms at 200\,Hz and
use only differences above 200\,Hz in the calculation, and also consider the
same finite parameter spacings. For small differences in EOS, the PN inspiral
models accurately mimic the measurability estimates of hybrid EOS. The
nonlinear behavior is also seen, but for large EOS differences there is some
overestimate from using inspiral-only models; this could be improved by using
post-Newtonian inspirals cut-off at a representative merger frequency.

\section{Multiple Signals}
A combination of $N$ identical signals, each with uncertainty $\delta\Lambda$ would give an
overall uncertainty $\delta\Lambda/\sqrt{N}$ if all events occurred at the same effective distance $D_\text{eff}$. When $\delta\Lambda$ scales
linearly with $D_\text{eff}$, we can use the results of \cite{Markakis2010} to
estimate how the uncertainty $\delta\Lambda_\text{0}$ at a reference
$D_\text{eff,0}$ translates to an expected combined error $\delta\Lambda$ from
signals randomly distributed within a horizon distance $D_\text{horizon}$. The combined uncertainty from the signals is given by
\begin{equation} 
\label{eq:combined}
\langle\delta\Lambda^{-2}\rangle^{-1/2} = \delta\Lambda_0
\frac{D_\text{horizon}}{D_\text{eff,0}}\left(3 N\right)^{-1/2}, \end{equation}
where $N$ is the total number of events.  

For the numerical-only estimates, linear scaling does not apply. 
We take the maximum range where we have calculated distinguishability of
signals, $\delta\Lambda\simeq2000$ at $D_\text{eff}=200$\,Mpc, within which we
expect to find $N_{200}$ equalling  9\% of the total number of signals (again
following \cite{Abadie:2010cf}). We conservatively take linear scaling from
this limiting distinguishability, and find
$\langle\delta\Lambda^{-2}\rangle^{-1/2} \simeq 670 \sqrt{3/N_{200}}$. 

For hybrid estimates, we can use the dotted line in
Fig.~\ref{fig:diffsnrplothyb}, which gives $\delta\Lambda\simeq350$ for
$D_\text{eff}=300$\,\mpc, to provide a roughly linear scaling within the
horizon distance of 445\,\mpc.  Eq.~(\ref{eq:combined}) then gives an
estimate for the expected measurement uncertainty of
$\langle\delta\Lambda^{-2}\rangle^{-1/2} \simeq 21 \sqrt{40/N}$. 

We expect, from studies with mixed binaries \cite{Lackey2013}, that
correlations with mass parameters will increase $\delta\Lambda$ by a factor of
$\sim3$.  Full bayesian parameter estimation using post-Newtonian waveforms
including tidal terms suggests that a combination of multiple signals can still 
used to distinguish between realistic EOS\cite{delPozzo2013}. However, the
statistical uncertainty would then be significantly smaller than the systematic errors
estimated in this work, which do not decrease with number of signals.
Uncertainty in waveform modeling would therefore limit our ability to measure
EOS parameters using binary neutron star systems.

\section{Conclusions}

It is now clear that tidal effects due to the finite size of neutron stars can
produce a detectable signature in gravitational signals that are likely to be
observed by ground-based gravitational wave detectors such as Advanced LIGO.  The observation of
these tidal effects presents the possibility of measuring neutron star
properties which in turn will constrain models for the neutron-star EOS.  In particular, using \emph{only} numerical simulations of the final
orbits, we estimate two EOS, H and B, which produce isolated neutron-star radii
that differ by $\delta R \sim 1.3\,\text{km}$, are distinguishable at
$D_\text{eff}=100$\,Mpc. This gives an effective $\delta R/R\sim10$\%. However,
the measurement accuracy does not improve linearly with SNR, and weaker signals
will have less discriminatory power. 

If trusted hybrids can be constructed, incorporating additional information
from the tidal post-Newtonian terms, then measurement errors drop
significantly, and the above EOS can be distinguished at
$D_\text{eff}=300$\,Mpc.  Numerical relativity efforts are required to generate
the waveforms needed to make these measurements, but the current
state-of-the-art simulations are already up to the task for large-radius
neutron stars. Future advances in numerical relativity will provide waveforms
of higher accuracy, and extending to lower frequencies, which will reduce
systematic errors on measurements of tidal effects.  Further improvements in
the scope of physical processes that are simulated by numerical relativity will
also enable us to follow the waveform through binary coalescence and past
merger, and therefore could allow for the measurement of additional EOS
properties from oscillations of a post-merger hypermassive remnant.

\acknowledgments

This work was supported by Grant-in-Aid for Scientific 
Research (24740163) of Japanese MEXT, by NSF awards
PHY-1055103,
PHY-0970074,
PHY-0900735,
PHY-0701817,
PHY-0503366, and PHY-1001515, and by the Grant-in-Aid for Young Scientists
(22740163). B.G. acknowledges support from NSF grant no. AST 1009396 and
NASA Grant No. NNX12AO67G. This work used XSEDE (allocation
TG-PHY110027) which is supported by NSF Grant No. OCI-1053575.  K.K.  is
supported by JSPS Postdoctoral Fellowship for Research Abroad.
CM acknowledges support by NSF Grant PHY1001515, DFG grant SFB/Transregio 7
``Gravitational Wave Astronomy" and STFC grant PP/E001025/1.
We thank 
B. D. Lackey for reading a draft of the paper and suggesting changes, 
for helpful conversations, and for providing Fig.~\ref{fig:eos-Lambda-R}. 
 J.S.R. thanks the the
``Rattle and Shine'' workshop at KITP (Santa Barbara) and the ``YKIS2013''
workshop at YITP (Kyoto) for useful discussions.

\bibliography{hybns}
\end{document}